\AtBeginDocument{%
	\paperwidth=\dimexpr
	1in + \oddsidemargin
	+ \textwidth
	+ 1in + \oddsidemargin
	\relax
	\paperheight=\dimexpr
	1in + \topmargin
	+ \headheight + \headsep
	+ \textheight
	+ 1in + \topmargin
	\relax
	\usepackage[pass]{geometry}\relax
}

\documentclass[envcountsame,envcountchap,natbib,a4paper,smallextended,twocolumn]{svjour3}

\usepackage{amsmath,xfrac}
\usepackage{overpic}
\usepackage{algorithm}
\usepackage{algpseudocode}

\DeclareMathAlphabet{\mathbbcal} {OMS}{cmsy}{m}{n}

\usepackage{mathrsfs}

\usepackage{amssymb} % math symbols
\usepackage{newtxmath}
\usepackage{type1cm}        % activate if the above 3 fonts are
\usepackage{makeidx}         % allows index generation
\usepackage{graphicx}        % standard LaTeX graphics tool
\usepackage{multicol}        % used for the two-column index
\usepackage[bottom]{footmisc}% places footnotes at page bottom
\usepackage{bm}

\usepackage{tcolorbox}
\usepackage{graphbox}
\usepackage{enumitem}
\usepackage{xcolor}
\usepackage[titletoc]{appendix}
\definecolor{black}{rgb}{0.08,0.08,0.08}
\color{black}   

\usepackage[hidelinks]{hyperref}

\usepackage{array,tabularx}
\usepackage[retainorgcmds]{IEEEtrantools}
\usepackage{etoolbox}
\usepackage{xcolor,tikz,pgfplots}
\definecolor{lightergray}{rgb}{0.9,0.9,0.9}

\usepackage{stmaryrd}

\usepackage{soul}

\usetikzlibrary{matrix,calc,positioning,decorations.markings,decorations.pathmorphing,decorations.pathreplacing}%tikzmark,
\usetikzlibrary{arrows,cd}
\usepackage{bbding}
\usetikzlibrary{positioning}
\tikzset{>=stealth}

\usepackage{datetime}
\usepackage{setspace}
\newdateformat{monthdayyeardate}{%
	\monthname[\THEMONTH]~\THEDAY, \THEYEAR}

\newdateformat{monthyeardate}{%
	\monthname[\THEMONTH], \THEYEAR}

\ifx \undefined \varvec    \def \varvec    #1{{\bm{#1}}} \fi

\RequirePackage{fix-cm}
\usepackage{xcolor}
\PassOptionsToPackage{dvipsnames,svgnames}{xcolor} 
\usepackage[object=vectorian]{pgfornament}

\newcommand{\R}{\mathbb R}

\newcommand{\N}{\mathbbcal{N}}

\newcommand{\norm}[1]{\left\lVert#1\right\rVert}

\def\bb{\mathbb }

\def\E{{\bb E}}
 
\def\E{{\bb E}}
\def\:{:\,}

  \def\ba{\begin{array}}
  \def\ea{\end{array}}
  \def\bc{\begin{corollary}}
  \def\ec{\end{corollary}}
  \def\bd{\begin{definition}}
  \def\ed{\end{definition}}
  \def\ben{\begin{enumerate}}
  \def\een{\end{enumerate}}
  \def\bse{\begin{equation*}}
  \def\ese{\end{equation*}}
  \def\be{\begin{example}}
  \def\ee{\end{example}}
  \def\bi{\begin{IEEEeqnarray*}}
  \def\ei{\end{IEEEeqnarray*}}
  \def\bit{\begin{itemize}}
  \def\eit{\end{itemize}}
  \def\bl{\begin{lemma}}
  \def\el{\end{lemma}}
  \def\bnn{\begin{notation}}
  \def\enn{\end{notation}}
  \def\bn{\begin{note}}
  \def\en{\end{note}}
  \def\bp{\begin{proposition}}
  \def\ep{\end{proposition}}
  \def\bq{\begin{proof}}
  \def\eq{\end{proof}}
  \def\br{\begin{remark}}
  \def\er{\end{remark}}
  \def\bs{\begin{solution}}
  \def\es{\end{solution}}
  \def\btab{\begin{table}}
  \def\etab{\end{table}}
  \def\btb{\begin{tabular}}
  \def\etb{\end{tabular}}
  \def\bt{\begin{theorem}}
  \def\et{\end{theorem}}

  \def\F{\mathbb{F}}

  \def\R{\mathbb{R}}

\def\sl{\mathfrak{sl}}

\DeclareFontFamily{U}{MnSymbolC}{}
\DeclareSymbolFont{MnSyC}{U}{MnSymbolC}{m}{n}
\DeclareMathSymbol{\diamondplus}{\mathbin}{MnSyC}{"7C}
\DeclareMathSymbol{\diamonddot}{\mathbin}{MnSyC}{"7E}
\DeclareFontShape{U}{MnSymbolC}{m}{n}{
    <-6>  MnSymbolC5
   <6-7>  MnSymbolC6
   <7-8>  MnSymbolC7
   <8-9>  MnSymbolC8
   <9-10> MnSymbolC9
  <10-12> MnSymbolC10
  <12->   MnSymbolC12}{}

\renewcommand{\F}{{F}}

\newcommand{\subsubsubsection}[1]{\paragraph{#1}\mbox{}\\}
\begin{document}
	
\title{A Riemannian Approach to Multivariate Geostatistical Modeling}
%\subtitle{Do you have a subtitle?\\ If so, write it here}

\titlerunning{A Riemannian Approach to Multivariate Geostatistical Modeling}        % if too long for running head

\author{\'Alvaro I. Riquelme*
}

%\authorrunning{Short form of author list} % if too long for running head

\twocolumn[
\begin{@twocolumnfalse}
	\maketitle
	\vspace{-3cm}
	\begin{abstract}
In geosciences, the use of classical Euclidean methods is unsuitable for treating and analyzing some
types of data, as this may not belong to a vector space. This is the case for correlation matrices, belonging to a subfamily of symmetric positive definite matrices, which in turn form a cone shape Riemannian manifold. We
propose two novel applications for dealing with the problem of accounting with the non-linear behavior usually
presented on multivariate geological data by exploiting the manifold features of correlations matrices. First, we employ an extension for the linear model of coregionalization (LMC) that alters the linear mixture, which is
assumed fixed on the domain, and making it locally varying according to the local strength in the dependency
of the coregionalized variables. The main challenge, once this relaxation on the LMC is assumed, is to solve
appropriately the interpolation of the different known correlation matrices throughout the domain, in a reliable and coherent fashion. The present work adopts the non-euclidean framework to achieve our objective by locally averaging and interpolating the correlations between the variables, retaining the intrinsic geometry of correlation matrices. A second application deals with the problem of clustering of multivariate data.

	\keywords{Geostatistical modeling \and Linear model of coregionalization \and Eigen-decomposition \and geodesics \and Riemannian manifold \and Symmetric positive definite}

	% \PACS{PACS code1 \and PACS code2 \and more}
	% \subclass{MSC code1 \and MSC code2 \and more}
	\end{abstract}

\end{@twocolumnfalse}
]

%	\institute{\'Alvaro I. Riquelme \at
%	Robert M. Buchan Department of Mining, Queen’s University, Kingston, Canada \\
%	\email{alvaro.riquelme@queensu.ca}           %  \\
%	%             \emph{Present address:} of F. Author  %  if needed
%	}  

{
	\renewcommand{\thefootnote}%
	{\fnsymbol{footnote}}
	\footnotetext[1]{\email{alvaro.riquelme@queensu.ca} \\
	Robert M. Buchan Department of Mining, Queen’s University, Kingston, Canada}
}

%\maketitle

%\begin{abstract}
%	
%
%\end{abstract}

\section{Introduction}
\label{intro}

%Motivation - need for accuracy and precision
In geosciences, classical Euclidean methods are not suitable for treating and analyzing some types of data, as they may not belong to a vector space. A common example is weather data, commonly assumed to be restricted to the sphere. Seeing the data as lying in different submanifolds of a Riemannian space is an increasingly used approach that has been highly successful over the past decades. In geostatistics, these concepts has been particularly used for the modeling of spatial non-stationarity in the data \citep{sampson1992nonparametric,almendral2008multidimensional,boisvert2009kriging,fouedjio2015estimation}. Other exhaustive application of the concepts have been done by \citeauthor{taylor2003euler} \citeyear{taylor2003euler}, \citeauthor{taylor2006gaussian} \citeyear{taylor2006gaussian} and \citeauthor{adler2007random} \citeyear{adler2007random} in order to understand the topology of random fields (RF) in manifolds. 

We address the problem of capturing and incorporating a second source of non-stationarity from the geological phenomena, which has to do with the fact that the different variables that describe ore deposits, $ Z_1(\textbf{u}),\dots$ , $Z_p(\textbf{u}) $, cannot be modeled  independently among them since they are mineralogically and physically related in complex fashions. As simple linear multivariate features rarely occur among geological variables composing sampling databases, usually showing nonlinear features instead, the correct reproduction of such characteristics becomes a problem when employing traditional estimation and geostatistical simulation techniques.

When relationships are simple and linear, one can rely on the \textit{linear model of coregionalization} (LMC) \citep{journel1978mining,Chiles} which can be interpreted, in the standard Gaussian setting, as assigning a constant correlation $ \rho_{ij} $   to the pair of variables $\{ Z_i(\textbf{u}),Z_j(\textbf{u})\} $ throughout the domain. This correlation parameter fixes in the space the direct and cross covariances theoretically, even at different positions $\{ Z_i(\textbf{u}),Z_j(\textbf{u}+\textbf{h})\} $, and must be modeled in beforehand to proceed with estimation techniques such as \textit{co-kriging} \citep{wackernagel2013multivariate}, or before applying decorrelation of the data through linear transformations such as Principal Component Analysis (PCA) \citep{pearson1901liii} or Minimum/Maximum Autocorrelation Factors \citep{switzer1985min}.

Since geological variables rarely show a linear Gaussian characteristics, it is hard to give to the LMC a geological interpretation, specially when a non-linear multivariate behavior among the different attributes is present, reducing the rate of success for the traditional methods. To overcome this limitation, some approaches that generalize the Gaussian transformation approach from the geological variables (\textit{raw} variables hereafter) into independent standard Gaussian variables, that can be treated individually, has been proposed \citep{leuangthong2003stepwise,barnett2014projection,van2017affine}. We follow a different path, which is to modify the LMC to alter the fixed correlation among geological features on the domain. This linear mixture can be made \textit{locally varying} according to the local strength in the dependency of the variables, leading to a \textit{locally varying linear model of coregionalization} (LVLMC), first introduced by \cite{gelfand2003spatial} in the context of spatial non-stationary models.

The main challenge, once the relaxation on the LMC is assumed, is to properly carrying out the correlation matrix, computed first at data position, to unknown locations of the spatial domain. Correlation matrices belong to the family of symmetric positive definite (SPD) matrices, which in turn forms a cone shape Riemannian manifold. Building upon earlier studies that have shown that a Riemannian framework is appropriate to address the challenge of interpolation between correlation matrices, the present work adopts this non-euclidean framework to achieve our objective by interpolating the correlations between the variables throughout the geological domain, retaining the intrinsic geometry of correlation matrices.

\section{Background}

\subsection{Review of Riemannian Manifolds}

\begin{figure}[h]
	\begin{center}
		\begin{overpic}[abs,unit=1mm,scale=1.]{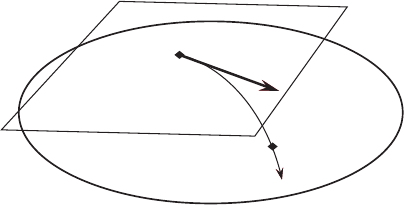}
			\put(9,9){\color{black}$ M $}
			\put(60,30){\color{black}$ T_{\textbf{x}_i}M $}
			\put(46,11){\color{black}$ \textmd{exp}_{\textbf{x}_i}(\textbf{v}) = \textbf{x}_j$}
			\put(47,20){\color{black}$ \textbf{v} $}
			\put(30,20){\color{black}$ \textbf{x}_i $}
		\end{overpic}
	\end{center}
	\caption{The exponential map.}
	\label{tange}
\end{figure}
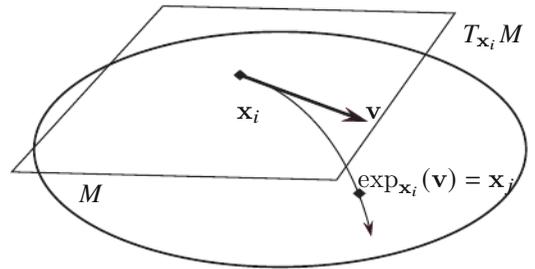

A differentiable manifold $ M $ of dimension $ p $ is a topological space  that is locally similar to a Euclidean space, with every point on the manifold having a neighborhood for which there exists a homeomorphism
(a continuous bijection whose inverse is also continuous) mapping the neighborhood to $ \R^{p} $. The tangent space $ T_\textbf{x}M $ at $ \textbf{x} $ is the vector space that contains the tangent vectors to all $ 1 $-D curves on $ M $ passing
through $ \textbf{x} $. Fig. \ref{tange} shows an example of a two-dimensional
manifold, a smooth surface living in $ \R^{3} $. A Riemannian
metric on a manifold $ M $ is a bilinear form which associates
to each point $ \textbf{x} \in M $ a differentiable varying inner product
$ \langle \cdot,\cdot \rangle^{}_{\!\textbf{x}} $ on the tangent space $ T_\textbf{x}M $ at $ \textbf{x} $. The norm of a vector
$ \textbf{v} \in T_\textbf{x}M $ is denoted by $ \norm{v}^2_\textbf{x} = \langle \textbf{v},\textbf{v} \rangle^{}_{\!\textbf{x}} $. The Riemannian
distance between two points $ \textbf{x}_i $ and $ \textbf{x}_j $ that lie on the manifold, $ d(\textbf{x}_i, \textbf{x}_j) $, is defined as the minimum length over all possible smooth curves on the manifold between $ \textbf{x}_i $ and $ \textbf{x}_j $. The smooth curve with minimum length is known as the
geodesic curve $ \gamma $.

Given a tangent vector $ \textbf{v} \in T_\textbf{x}M $, locally there exists a
unique geodesic $ \gamma_\textbf{v}(t) $ starting at $ \textbf{x} $ with initial velocity $ \textbf{v} $,
and this geodesic has constant speed equal to $ \norm{v}^2_\textbf{x} $. The
exponential map, $ \textmd{exp}_\textbf{x}: T_\textbf{x}M  \rightarrow M$ maps a tangent vector $ \textbf{v} $ to the point on the manifold that is reached at time $ 1 $ by the geodesic $ \gamma_\textbf{v}(t) $. The inverse of $ \textmd{exp}_\textbf{x} $ is known as the logarithm map and is denoted by $ \textmd{log}_\textbf{x}: M  \rightarrow T_\textbf{x}M$. Now, if we have two points $ \textbf{x}_i $ and $ \textbf{x}_j $ on the manifold $ M $, the tangent vector to the geodesic curve from $ \textbf{x}_i $ to $ \textbf{x}_j $ is defined as $ v = \textmd{log}_{\textbf{x}_i}(\textbf{x}_j) $, and the exponential map takes $ \textbf{v} $ to the point $ \textbf{x}_j  = \textmd{exp}_{\textbf{x}_i}\big(\textmd{log}_{\textbf{x}_i}(\textbf{x}_j)\big)$. In addition, $ \gamma_\textbf{v}(0) = \textbf{x}_i $
and $ \gamma_\textbf{v}(1) = \textbf{x}_j $. The Riemannian distance between $ \textbf{x}_i $ and $ \textbf{x}_j $ is defined as $ d(\textbf{x}_i, \textbf{x}_j)=\norm{\textmd{log}_{\textbf{x}_i}(\textbf{x}_j)}_{\textbf{x}_i} $.

Given the data $  \textbf{x}_1 ,\dots,\textbf{x}_n \in M $, we consider the use  \textit{\textbf{geometric}} or \textbf{\textit{Fr\'echet mean}} $ \varvec\mu $ is defined as a minimizer of the sum of squared distances: $$\varvec\mu = \textmd{arg} \inf_{\substack{\textbf{x}} \in M} \sum_{i=1}^k d^2(\textbf{x},\textbf{x}_i).$$

\subsection{The Riemannian manifold of SPD matrices}

Let $ \textrm{Sym}^+(p) $
denote the set of symmetric, positive definite matrices of size $ p \times p $, that is the set of all symmetric $ p \times p $ matrices $ \textbf{X} $ such that the quadratic form
$ \textbf{v}^T\textbf{X}\textbf{v} > 0 \textrm{ for all } \textbf{v} \in \R^p $. A crucial aspect of the
set $ \textrm{Sym}^+(p) $ is that it is not a vector space but forms a cone-shape Riemannian manifold. As a consequence of the manifold
structure of $ \textrm{Sym}^+(p) $, computational methods that simply rely
on the Euclidean distances between SPD matrices are generally sub optimal, with low performance \citep{4479482}. It is necessary to consider the notion of geodesic distance to exploit the manifold structure of $ \textrm{Sym}^+(p) $, which is the length of the shortest curve connecting two points, in this case two matrices, on the manifold.
Among the different Riemannian metrics that have been
considered on $ \textrm{Sym}^+(p) $, the one that has been most studied and
analyzed is the classical \textit{affine-invariant metric}, in which the
geodesic distance on the manifold between two SPD matrices
$ \textbf{P}_1 $ and $\textbf{ P}_2 $ is defined as:
\begin{eqnarray} d_{\textrm{Sym}^+}^2(\textbf{P}_1, \textbf{P}_2)&=&  \norm{\text{Log}(\textbf{P}^{1/2}_1\textbf{P}^{1/2}_2\textbf{P}^{1/2}_1)}^2\nonumber\\ &=& \text{tr}\big(\text{Log}^2(\textbf{P}^{1/2}_1\textbf{P}^{1/2}_2\textbf{P}^{1/2}_1)\big) \nonumber
\end{eqnarray}
with $ \text{Log}(\cdot) $ denoting the matrix logarithm and $ \norm{\textbf{A}}^2 = \text{tr}(\textbf{A}^{T}\textbf{A}) $ denoting the Frobeniuous matrix norm.

Furthermore, given a tangent vector $ \textbf{Y}_\textbf{P} \in T_\textbf{P}\textrm{Sym}^+(p)$ at a point $ \textbf{P} \in \textrm{Sym}^+(p) $, the Riemannian exponential map $\textmd{exp}_{\textbf{P}}: T_{\textbf{P}}M \rightarrow \textrm{Sym}^+(p)$ is given by 
\begin{equation}
	\textbf{X}=\textmd{exp}_{\textbf{P}}(\textbf{Y}_\textbf{P})=\textbf{P}^{1/2}\textmd{Exp}(\textbf{P}^{-1/2}\textbf{Y}_\textbf{P}\textbf{P}^{-1/2})\textbf{P}^{1/2}\text{.}	\label{exp}
\end{equation}
where $ \text{Exp}(\cdot) $ denotes the exponential of a matrix. Given two positive definite matrices $\textbf{P}, \textbf{X} \in\textrm{Sym}^+(p)$, the Riemannian logarithmic map $\textmd{log}_{\textbf{P}}: \textrm{Sym}^+(p) \rightarrow T_{\textbf{P}}\textrm{Sym}^+(p)$, of $ \textbf{X}$  in relation to $ \textbf{P}$ is given by
\begin{equation}
	\textbf{Y}_\textbf{P}=\textmd{log}_{\textbf{P}}(\textbf{X})=\textbf{P}^{1/2}\textmd{Log}(\textbf{P}^{-1/2}\textbf{X}\textbf{P}^{-1/2})\textbf{P}^{1/2}\text{.}\label{logmap}
\end{equation}

Finally, the geodesic passing through \textbf{P} in the direction of $ \textbf{Y}_\textbf{P}$ is uniquely given by
\begin{equation}
	\gamma_\textbf{P}(t;\textbf{Y}_\textbf{P}) =  {\textbf{P}^{1/2}}\textmd{Exp}( {\textbf{P}^{-1/2}}\textbf{Y}_\textbf{P} {\textbf{P}^{-1/2}}t) {\textbf{P}^{1/2}}\text{.}	\label{geod}
\end{equation}

\subsection{The Riemannian manifold of Correlation matrices}

\subsubsection{Visualizing $ \textrm{Corr}(2) $ and $ \textrm{Corr}(3) $}
The affine-invariant structure for $ \textrm{Sym}^+(p) $ is not only intrinsically linked with $ \textrm{Corr}(n) $ but also imposes  symmetry on its structure as a quotient manifold.

\begin{figure}[h!]
	\centering
	\includegraphics[width=0.45\textwidth,trim={0 0cm 0.0cm 0cm},clip]{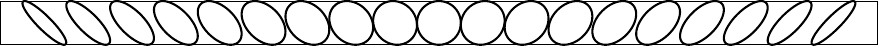}
	\caption{The manifold $\textrm{Corr}(2) $.}
	\label{corrman}
\end{figure}

\begin{figure}
	\centering
	\includegraphics[width=0.45\textwidth,trim={0 0cm 0.0cm 0cm},clip]{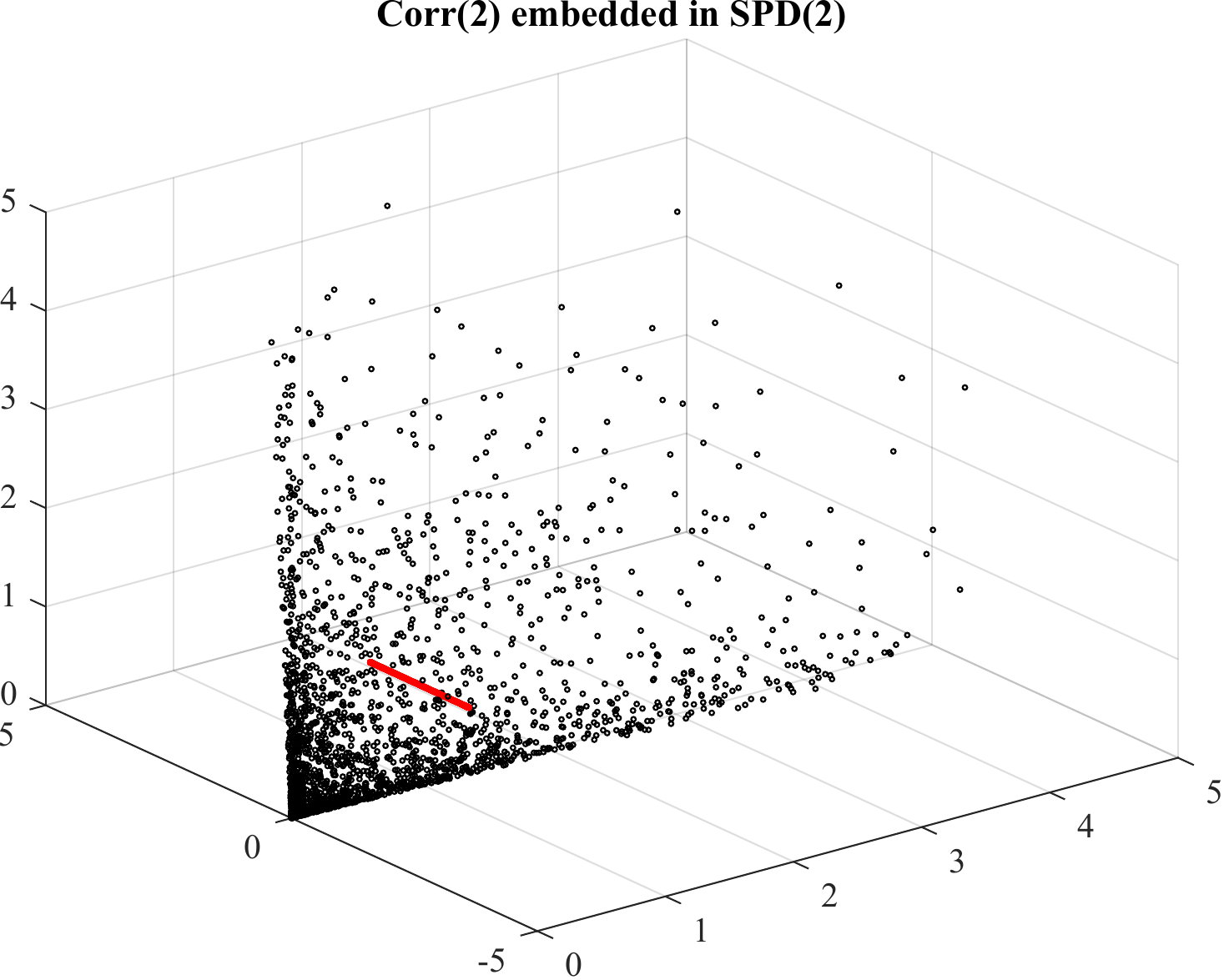}
	\caption{The manifold $\textrm{Corr}(2) $ visualized as an embedded submanifold of $ \textrm{Sym}^+(2) $. Points in $ \textrm{Sym}^+(2) $ (black) are sampled independently of those in $\textrm{Corr}(2) $	(red).}
	\label{corrman23}
\end{figure}

Let us begin by visualizing $ \textrm{Corr}(2) $ as a subset of $ \textrm{Sym}^+(2) $:
\begin{equation*}		
	\textrm{Corr}(2):=\Biggl\{
	\begin{pmatrix}
		1 &\quad x\\
		x &\quad 1\\
	\end{pmatrix} \quad : \quad x \in (-1,1)\Biggr\}\text{.}
\end{equation*}
then we can realize this space to be a manifold of dimension 1 parameterized by the map $ \varphi : (-1, 1) \rightarrow \textrm{Corr}(2) $ given by
\begin{equation*}		
	\varphi(x) = 
	\begin{pmatrix}
		1 &\quad x\\
		x &\quad 1\\
	\end{pmatrix}\text{.}
\end{equation*}
This is a smooth map into the symmetric matrices which restricts to $ \textrm{Corr}(2) $ whose inverse is simply given by projection onto one of the off-diagonal entries. We can visualize any $ \textbf{C} \in \textrm{Corr}(2) $ by associating the ellipsoid $ \textbf{v}^T\textbf{C}^{-1}\textbf{v} =1$. Because of the global parametrization $ \varphi : (-1, 1) \rightarrow \textrm{Corr}(2) $, we can visualize the manifold $\textrm{Corr}(2) $ as the interval $ (-1, 1) $, but at each point in the interval we can attach to it the
ellipsoid corresponding to the positive-definite form associated to the matrix. We see this in Fig. \ref{corrman}. Another visualization we will consider is to see the correlation
matrices embedded inside the symmetric positive-definite matrices (Fig. \ref{corrman23}).

\begin{figure}
	\centering
	\includegraphics[width=0.325\textwidth,trim={0 0cm 0.0cm 0cm},clip]{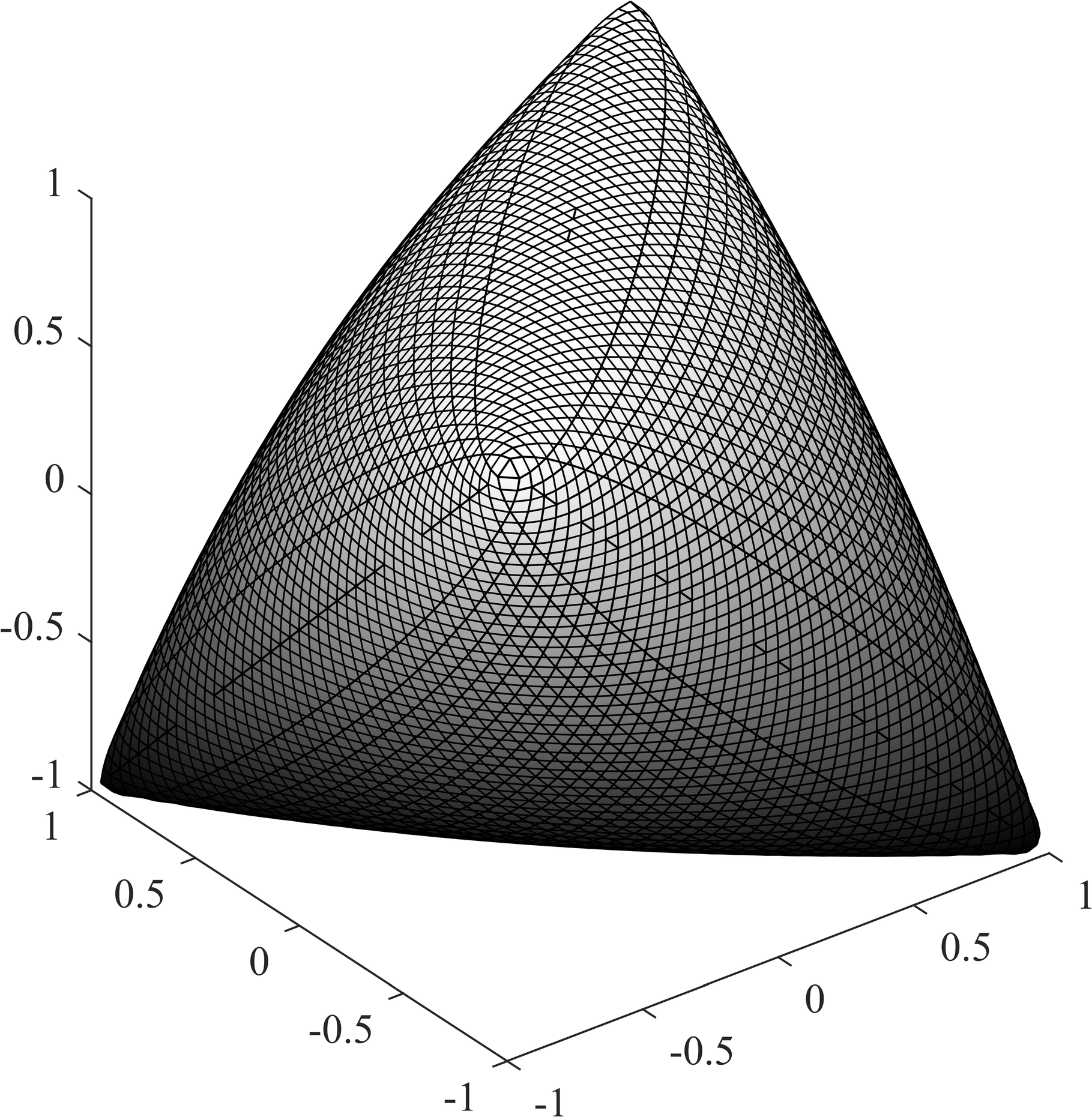}%
	\caption{The boundary of the elliptope.}
	\label{elip}
\end{figure}

In the case of correlation matrices of dimension $ 3 $, the shape formed by the set is named the 3-dimensional \textbf{\textit{elliptope}} which can be represented as a linear matrix inequality, characterize by

\begin{equation*}		
	\textrm{Corr}(3):=\Biggl\{
	\begin{pmatrix}
		1 &\quad x&\quad y\\
		x &\quad 1&\quad z\\
		y &\quad z&\quad 1\\
	\end{pmatrix} \quad : \det \begin{bmatrix} 1 &\quad x &\quad y\\ x &\quad 1 &\quad z\\ y &\quad z &\quad 1\end{bmatrix} = 
\end{equation*}
\[ 1 + 2 x y z - x^2 - y^2 - z^2 > 0\Biggr\}\text{.} \]
The boundary of the elliptope (Fig. \ref{elip}) is the cubic surface defined by
\begin{equation*}		
1 + 2 x y z - x^2 - y^2 - z^2 = 0
\end{equation*}

\subsubsection{Quotient Geometry}

\begin{figure}
	\centering
	\includegraphics[width=0.45\textwidth,trim={0 0cm 0.0cm 0cm},clip]{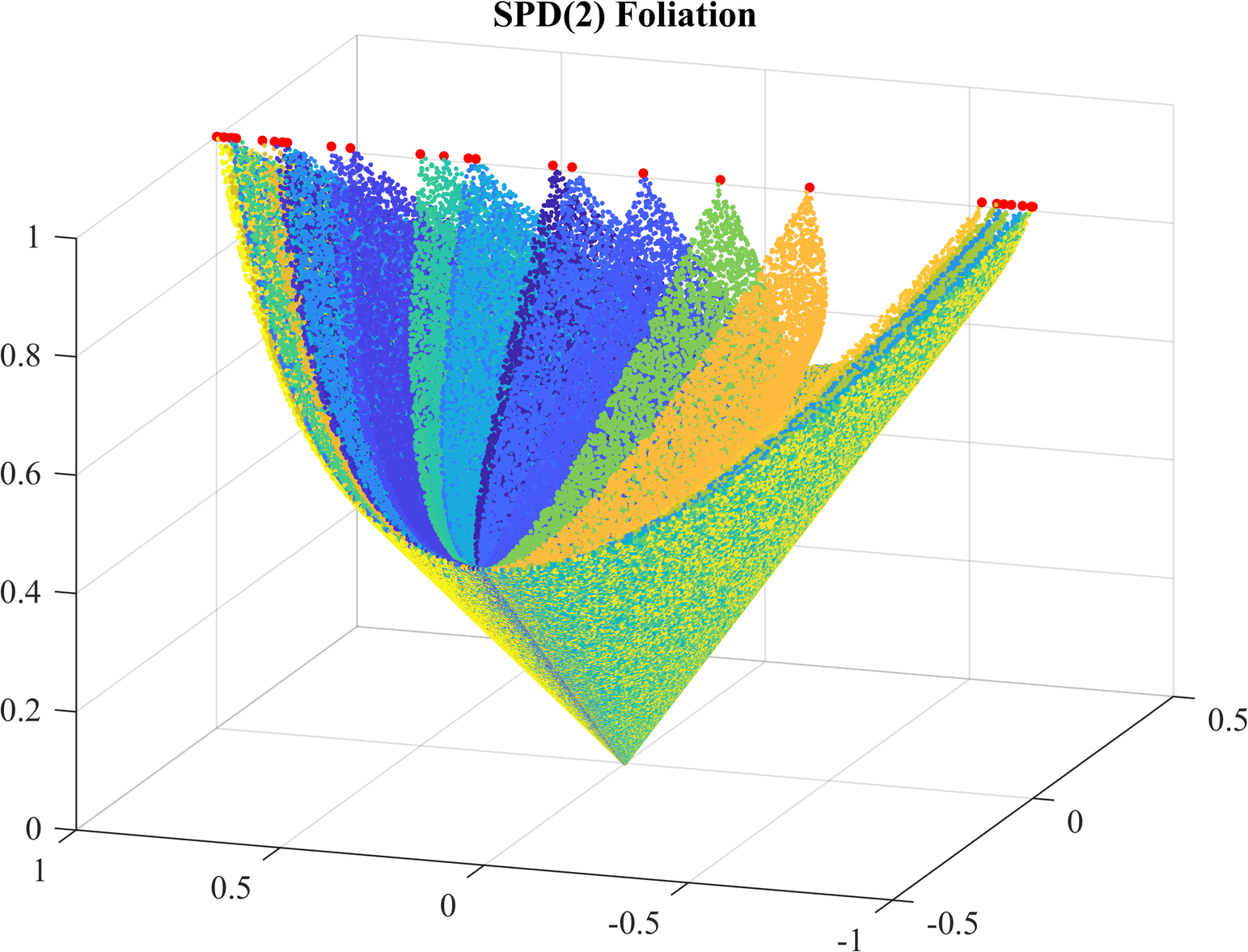}%
	\caption{A foliation of the cone $\textrm{Sym}^+(2) $. Each leaf is an embedded two-dimensional submanifolds, obtained by translating a given correlation matrix $ \textbf{C} $ (red dot) by the action $ \textbf{DCD} $.}
	\label{fol}
\end{figure}

Consider an element ${\varvec{\upSigma}} \in \textrm{Sym}^+(p) $. The orbit of ${\varvec{\upSigma}}$, that is, the set of images of ${\varvec{\upSigma}}$ when considering the action of a group diagonal matrices with positive entries $ \textmd{Diag}^{+}(p) $ on it, $\textmd{Diag}^{+}(p) \times \textrm{Sym}^+(p) \rightarrow \textrm{Sym}^+(p)$, given by $ (\textbf{D},{\varvec{\upSigma}}) \mapsto \textbf{D}{\varvec{\upSigma}}\textbf{D} $:
\[ \textmd{Diag}^{+}(p)\, \varvec\cdot\, \varvec{\upSigma} = \big\{\textbf{D}\,\varvec\cdot\, \varvec{\upSigma} : \textbf{D} \in  \textmd{Diag}^{+}(p)\big\}\text{,} \qquad \textbf{D}\,\varvec\cdot\, \varvec{\upSigma}:=  \textbf{D}{\varvec{\upSigma}}\textbf{D}\text{,}\]
is a smooth manifold of dimension equal to $ \textrm{dim Diag}^+(p) = p $. This can be seen explicitly in the case of taking an element $ \textbf{C} \in \textrm{Corr}(2) \subset \textrm{Sym}^+(2) $, and sampling the orbit space by applying $ \textbf{DCD} $, where $ D \in \textrm{Diag}^+(p) $ is generated randomly (Fig. \ref{fol}).

Subsequently the quotient manifold $\textrm{Sym}^+(p)/\textmd{Diag}^{+}(p) $ is a smooth manifold on which one can take, as representative of the equivalence relation, an element of $ \textrm{Corr}(p) $, with  $ \textrm{dim Corr}(p) =\textrm{dim Sym}^+(p)-\textrm{dim Diag}^+(p)$  \citep{david2019riemannian}. Intuitively, this correspond to a ``retraction'' along the leaves to the one dimensional line $\textrm{Corr}(2)$, for the case of $ \textrm{Sym}^+(2) $.

The representative that we take on $\textrm{Sym}^+(p)/\textmd{Diag}^{+}(p) $ correspond to the element given by the projection\[ \pi : \textrm{Sym}^+(p) \rightarrow \textrm{Corr}(p) \qquad  \varvec{\upSigma}\mapsto(\textbf{D}_{\varvec{\upSigma}},\varvec{\upSigma}) = \textbf{C}_{\varvec{\upSigma}}\text{,}\] $ $ where $ \textbf{D}_{\varvec{\upSigma}} = (\textbf{I}_p \circ \varvec{\upSigma})^{-1/2} $ and $ \circ $ is the Hadamard product. Since more than one element can be projected into the same correaltion matrix, we call to the leave $ \pi^{-1}(\textbf{C}_{\varvec{\upSigma}}) $ projected into the correlation matrix $ \textbf{C}_{\varvec{\upSigma}} $ the \textit{fiber} of $\textbf{C}_{\varvec{\upSigma}} $:
\[ \pi^{-1}(\textbf{C}_{\varvec{\upSigma}}) = \{\varvec{\upSigma} \in \textrm{Sym}^+(p) : (\textbf{D}_{\varvec{\upSigma}},\varvec{\upSigma}) = \textbf{C}_{\varvec{\upSigma}}\}\text{.} \]

\subsubsection{Accounting for a distance in $\textrm{Corr}(p)$}

While the result that $\textrm{Corr}(p)$ exhibits this particular quotient manifold structure is meaningful in itself, this fact alone does not yield results that are suitable for algorithms and computation as closed form expression are not available for computing distances on $\textrm{Corr}(p)$. Thus, one must rely only on Riemannian structure that $\textrm{Corr}(p)$ inherits from $ \textrm{Sym}^+(p) $ in order to obtain an algorithm that computes distances through an optimization procedure.

In order to come up with such an algorithm, it is used a really helpful theorem, proved by \cite{huckemann2010intrinsic}, showing that the geodesic connecting two points in the quotient can be expressed as the geodesic in the ambient manifold from the starting point to an
optimal representative of the end point, lying on the fiber over the desired endpoint:

\bt
(Huckemann 2010). Let $ M $ be a Riemannian manifold with an isometric
action of a Lie group $ G $. Then a geodesic $ \gamma $ in the quotient $ M/G $ with end points $ a, b \in M/G $ can be obtained from the projection of a geodesic $ \tilde\gamma $ on M (i.e. $ \gamma = \pi \circ \tilde\gamma$) such that
\begin{itemize}
	\item $\tilde\gamma $ has end points $ p$, $ q $ with $ \pi(p) = a $, $ \pi(q) = b $, and
	\item $ q $ is the solution to the problem \[ \text{min}\, d_M(p,c) \quad\text{such that } c\in  \pi^{-1}(b)\text{.}\]
\end{itemize}
This last point can be rephrased for fixed $ c \in  \pi^{-1}(b) $ as \[ \text{min}\, d_M(p,g \cdot c) \quad\text{such that } g\in  G\text{.}\] 
\et

Adapting equations \ref{geod} to the current scenario, let $\textbf{C}_1,\textbf{C}_2 \in \text{Corr}(n). $ Then the geodesic and corresponding distance in $ \textrm{Sym}^+(p) $ connecting these two points are given by the following:

\begin{eqnarray*}
	\gamma_{\textrm{Sym}^+}(t) = \textbf{C}_1^{1/2}\textmd{Exp}\big(\textmd{Log}(\textbf{C}_1^{-1/2}{\textbf{C}}_2\textbf{C}_1^{-1/2})t\big)\textbf{C}_1^{1/2}\text{,} \\
	d^2_{\textrm{Sym}^+}=\big|\big|\textmd{Log}(\textbf{C}_1^{-1/2}{\textbf{C}}_2\textbf{C}_1^{-1/2})\big|\big|^2 \text{.}
\end{eqnarray*} 

In order to adapt this Riemannian structure to $\textrm{Corr}(p)$ we need to find the optimal
representative of $ \textbf{C}_2 $ with respect to the starting point $ \textbf{C}_1 $. This is done by finding the unique element $ \widetilde{\textbf{C}}_2 $ in the fiber $ \pi^{-1}(\textbf{C}_2) $ which minimizes the $ {\textrm{Sym}^+}$-distance between $ \textbf{C}_1 $
and $ \widetilde{\textbf{C}}_2 $. This can be written as

$$ d^2_{\textmd{Corr}}(\textbf{C}_1,\textbf{C}_2)= \inf_{\substack{\textbf{D} \in \textmd{Diag}^{+}(p)}}d^2_{\textrm{Sym}^+}(\textbf{C}_1,\textbf{DC}_2\textbf{D})$$
Using this equation above we then aim to solve the following minimization problem:
\begin{equation}
	\text{minimize}\,d^2_{\textrm{Sym}^+}(\textbf{C}_1,\textbf{DC}_2\textbf{D})\quad\text{subject to}\quad\textbf{D}\in\textmd{Diag}^{+}(p)\text{.}	\label{mincorr}
\end{equation}
Assuming $ \textbf{D}^*$  is a sufficient solution to the above problem, we define as $ \widetilde{\textbf{C}}_2 $ this element in the fiber $ \pi^{-1}(\textbf{C}_2) $ which minimizes the $ {\textrm{Sym}^+}$-distance between $ \textbf{C}_1 $ and $\widetilde{\textbf{C}}_2$
$${\widetilde{\textbf{C}}}_2 = \textbf{D}^*\textbf{C}_2\textbf{D}^*\text{.}$$
The corresponding geodesic can be taken as the projection of the $ {\textrm{Sym}^+}$-geodesic connecting $ \textbf{C}_1 $ and and $ \widetilde{\textbf{C}}_2$  
$$ \gamma_{\textmd{Corr}}(t) = \pi \Big(\textbf{C}_1^{1/2}\textmd{Exp}\big(t\textmd{Log}(\textbf{C}_1^{-1/2}\widetilde{\textbf{C}}_2\textbf{C}_1^{-1/2})\big) \Big)\text{.}$$

\section{Applications}
\label{Applications}

On this section we present two novel applications based on the theory introduced previously. The first one has to do with the prediction of geological attributes at unknown locations, showing complex non-linear multivariate features on the data. The second one is related to clustering of data.

\subsection{Extending The Linear Model of Coregionalization}

As we mentioned earlier, the main contribution of this work is to see any geological process globally as a mixture of multivariate RVs on a given spatial domain $  D $, acting locally with different properties that change smoothly throughout the different positions $ \textbf{u} \in D $. The correlation among attributes is the property that we consider as a function of $ \textbf{u} $, as we consider standard Gaussian RVs, given as a result the reproduction of the complex non-linear features among variables. Therefore, this idea is a simple, linear, and geological meaningful approach to the estimation and uncertainty quantification at unknown locations when the mentioned characteristics are exhibited on the data.

\subsubsection{The Model}

The following model relies on the assumptions that ``simple'' non-linear multivariate features can be reconstructed in a straightforward way by mapping the original $ p $-variate cumulative distribution function with a $ p $-variate Gaussian distribution equipped with a proper prior covariance matrix. This procedure is also known as Nataf transformation \citep{nataf1962determination} or NORTA (NORmal To All), and several properties of the transformation has been studied in different contexts, for instance, in \cite{cario1997modeling,ayadi2019norta,xie2015quantifying,xiao2014evaluating,li1975generation} and on \cite{bourgault2014revisiting} in the geostatistical context. We start by a brief motivation proceeded by highlighting the relevant theoretical aspects of the transformation.

\begin{figure}[h]
	\begin{center}
		\includegraphics[width=0.48\textwidth]{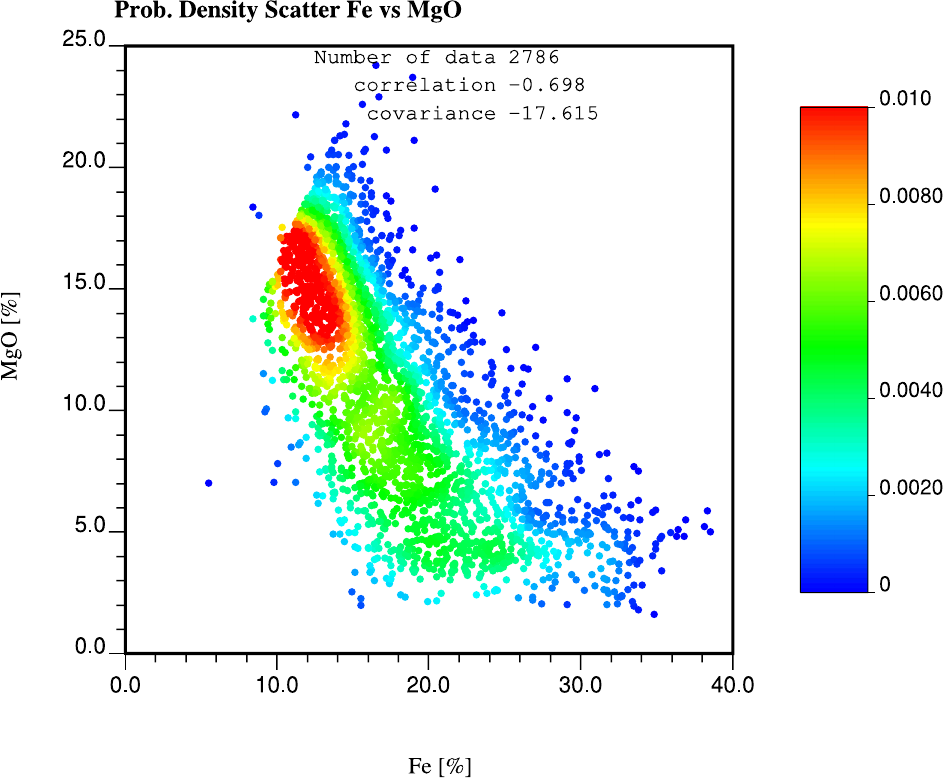}
		\includegraphics[width=0.48\textwidth]{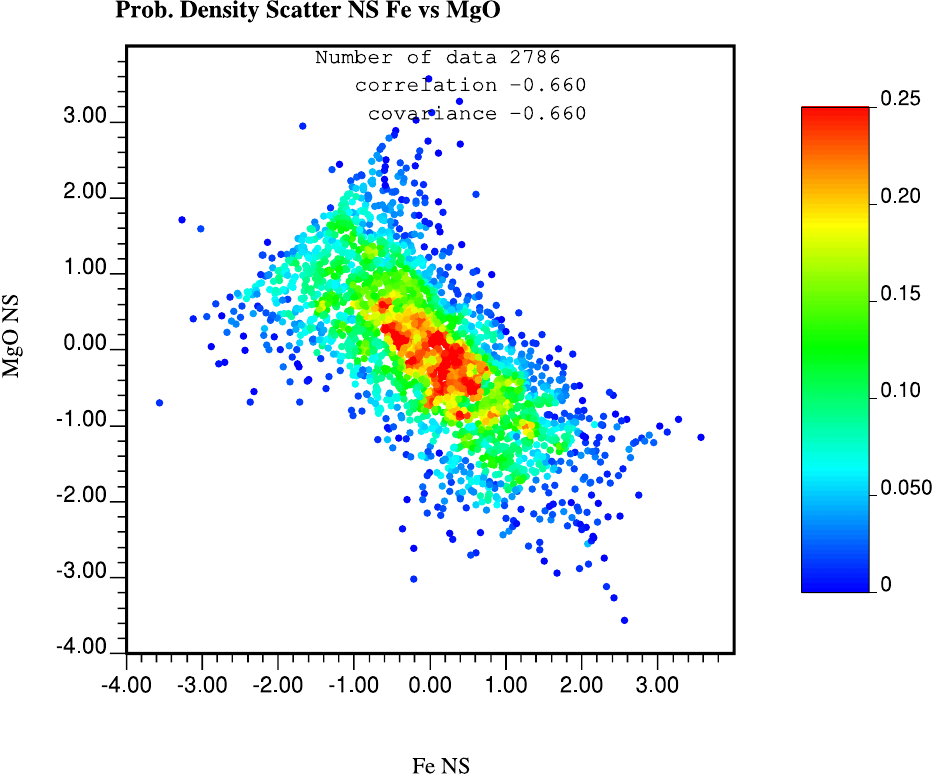}
	\end{center}
	\caption{Correlation of variables before and after applying a normal score transformation. Variables are still correlated.}
	\label{anamor}
\end{figure}

Let $ {{\tilde{\textbf{Z}}}}= [\tilde{Z}_1(\textbf{u}),\dots, \tilde{Z}_p(\textbf{u})]^T $ be the vector-valued random function (RF) considering $p$ simultaneous RFs  $\tilde{Z}_i=\{\tilde{Z}_i(\mathbf{u}): \mathbf{u} \in D \subseteq \R^n, n\geq 1\}$, indexed by $ i $ ranging in the set  $I=\{1, ... , p\}$, and defined on a fixed continuous domain of interest $D$ of the Euclidean space $\R^n$. Let the sampling data given by the multivariate vectors $ {{\tilde{\textbf{z}}}}_\alpha = [\tilde{z}_1(\textbf{u}_\alpha),\dots,\tilde{z}_p(\textbf{u}_\alpha)]^T $, $ \alpha \in \{1,\dots,k\} $, defined as \textit{data}. We face, as a main problem, that the blindly procedure the values $ \tilde{z}_i $ of the different RV $ \tilde{Z}_i $, transforming each variable into a uni-variate Gaussian values $ z_i $, \[ z_i=G^{-1}\big(F(\tilde{z}_i)\big), \] does not translate into independent Gaussian variables, $ Z_i $. This is shown in the cross plots of Fig. \ref{anamor}, giving a comparative illustration of the original data, normally transformed data in a uni-variate way, showing that after the transformation, the data, that was previously correlated in raw values, is still correlated after the transformation.

Therefore, when modeling two or more variables by using a non-correlated multi-Gaussian pdf, the path is prone to give bad results when estimating or simulating values when back transforming into raw values, since the separate transformation entails an incorrect map on the multivariate probability densities. However, this problem is quickly fixed when a correlated Gaussian distribution is considered instead.
This simple method works as a multivariate transformation, by coupling the univariate transformations $ \phi_i $.

We define the \textit{non-coupled} transformation of the initial multivariate RF ${{\tilde{\textbf{Z}}}}$ into a \textit{stationary} $p$-variate Gaussian RF with zero vector mean ${\varvec\mu}= (0,\dots,0)^T=\textbf{0} $ and covariance matrix equal to the identity matrix $\textbf{I}_p$, i.e., $\textbf{Z}=[Z_1(\textbf{u}),\dots,Z_p(\textbf{u})]^T \sim \N(\textbf{0},\textbf{I}_p)$ (${\textbf{I}_p}_{ii} = 1 \textmd{ and }{{\textbf{I}_p}_{ij} = 0}$, $i,j \in I$), by using the anamorphosis function $\phi_i^{-1}$ on each of the components of $Z$:
\begin{eqnarray}
	{{\tilde{\textbf{Z}}}}=&[\tilde{Z}_1(\textbf{u}),\dots,\tilde{Z}_p(\textbf{u})]^T&\nonumber\\
	=&(\phi_1[Z(\textbf{u})],\dots,\phi_p[Z(\textbf{u})])^T&={\Phi_\textbf{I}}_p(\textbf{Z})
	\text{.}\nonumber
\end{eqnarray}

The \textit{coupled} prior distribution of $\textbf{Z}$, is still a $p$-variate Gaussian distribution ${\N}( \textbf{0},\widehat{\varvec{\upSigma}})$, with mean vector and \textbf{\textit{correlation}} matrix given by
\begin{equation*}
	\widehat{\varvec{\upSigma}} = \begin{pmatrix}
		1&\quad\hat\rho_{12}&\quad\cdots&\quad\hat\rho_{1p} \\
		\hat\rho_{21}&\quad1&\quad\cdots&\quad\hat\rho_{2p} \\
		\vdots    &\quad\vdots      &\quad\ddots&\quad\vdots \\
		\hat\rho_{p1}&\quad\hat\rho_{p2}&\quad\cdots&\quad1
	\end{pmatrix}.
\end{equation*}

Then, the random variables  $Z_i$ are correlated and their pairwise relationships are quantified by the correlation coefficients $\hat\rho_{Z_iZ_j}$ (or simply $\hat\rho_{ij}$) with $i,j \in I$, which has to be inferred. We proceed to do this in the next section. The $p$-variate ccdf over the original variables is then retrieved simply as:
\begin{equation}
	\F_{\tilde{Z}_1(\textbf{u}),\dots,\tilde{Z}_p(\textbf{u})}(\tilde{z}_1,\dots,\tilde{z}_p)=G_\mathbf{0}^{\widehat{\varvec{\upSigma}}}\big(\phi_1^{-1}(\tilde{z}_1),\dots,\phi_p^{-1}(\tilde{z}_p)\big)\label{transnat}
\end{equation}

\begin{figure}[h]
	\begin{center}
		\includegraphics[width=0.49\textwidth]{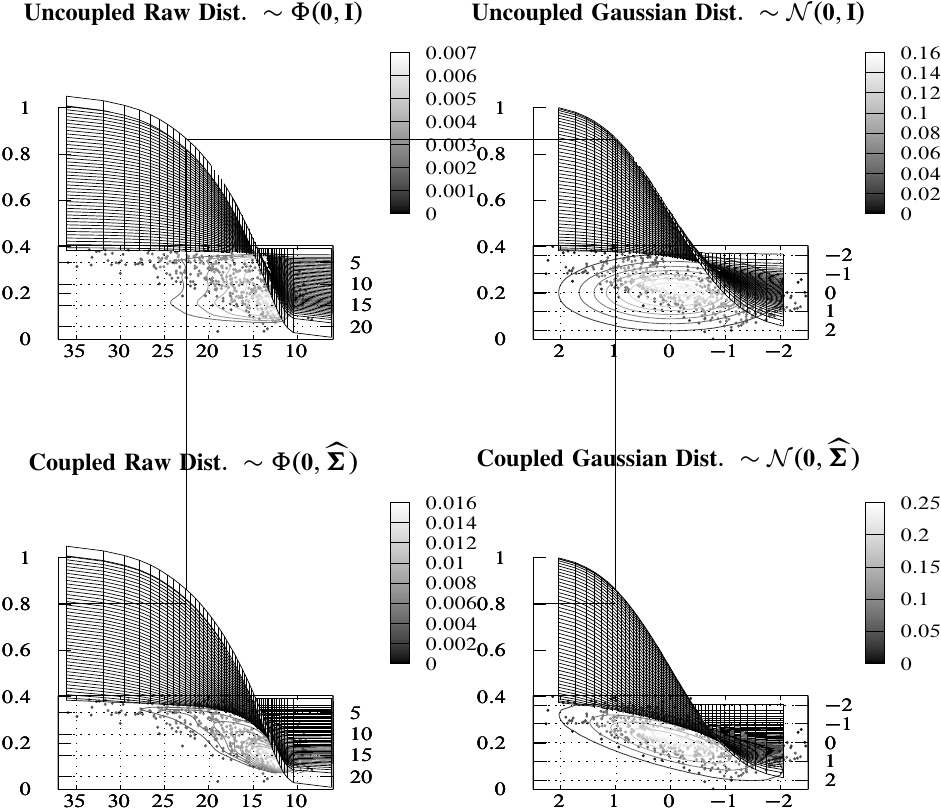}
	\end{center}
	\caption{Conceptual bi-variate picture of the adjustment in the correlated behavior of independent raw distribution though the Gaussian coupled anamorphosis. In surface are the correspondent cdfs, and in contour plots the pdfs.}
	\label{GAn3}
\end{figure}

We will say that $\textbf{Z}$ follows a \textit{coupled anamorphosis function}, i.e., $Z \sim \varvec\Phi( {\textbf{0}},{\widehat{\varvec{\upSigma}}})$. The transformation (or coupling process) is conceptually illustrated, for the bi-variate case, in Figure \ref{GAn3}. 

It is important to mention that this transformation is \textit{well-defined}, in the sense that the order of variables does not play a role, and a permutation of them just translates in permutation of the correlation coefficients on $ {\widehat{\varvec{\upSigma}}} $. However, this procedure entails the severe hypothesis that the multivariate behavior of geological attributes can be modeled by assuming a correlated Gaussian distribution, which may be a lousy model globally. Instead, we take this hypothesis for granted \textbf{\textit{locally}} in the geological domain.

Given the different RVs that describe ore deposits, $ {{\tilde{\textbf{Z}}}}= [\tilde{Z}_1(\textbf{u}),\dots, \tilde{Z}_p(\textbf{u})]^T $, we proceed to transform the variables into Gaussian RVs jointly, according to Eq. \ref{transnat}, in order to get the vector ${{{\textbf{Z}}}}=[{Z}_1(\textbf{u}_\alpha),\dots,{Z}_p(\textbf{u}_\alpha)]$. 

Once we have device to perform this transformation, we have to decide among two possible modeling options. On the first hand, to perform the gaussianization in a ``global'' fashion, that is, gathering all the $ \textbf{z}_\alpha $ data and perform only one transformation by running Eq. \ref{transnat} once. On the second hand, to perform the gaussianization ``locally'', which means to collect chunks of data in a vicinity $ V $ to the location under study, $ {{\tilde{\textbf{z}}}}_\alpha $, $ \alpha \in V \subset \{1,\dots,k\} $ in a moving neighborhood fashion, noticing that $ V $ may be a function of the location under study, $ V(\textbf{u}) $. This is a non-trivial choice to do. We take the second path in our model, since there is no loss of generality and contains the case on which $ {{\tilde{\textbf{Z}}}} $ is stationary, as performing gaussianization locally should not be theoretically biased. As the next step in the methodology is the inference of a local correlation matrix $\varvec\upSigma(\textbf{u}_\alpha) $ at the sampling locations, taking the first path of performing a global transformation and later taking chunks of data would give, as result, the inference of a $\varvec\upSigma(\textbf{u}_\alpha) $ matrix based on data with non-zero mean locally. Finally, the second path does not contradict the traditional methodology for uncertainty modeling, which consists in partitioning the data in stationary clusters, and continue the work individually on each of the clusters separately.

%%%%%%%%%%%%%%%%%%%%%%%%%%%%%%%%%%%%%%%%%
Then, the LMC is brought into play, and assuming that each variable consists of a sum of $ p $ independent factors:
\begin{equation}
	Z_i(\mathbf{u}) = \sum_{j=1}^{p}a_{ij}(\mathbf{u})Y_j(\mathbf{u}),
\end{equation}
with the number of factors equal to the number of attributes in order to avoid ill-definition as a linear system (the problem of working in the stationary set-up with a number of factors different to the number of attributes has been recently tackled by \cite{pinto2021decomposition}, and including their methodology into the presented one is a topic of further research). Thus, obtaining a model in the fashion $\textbf{Z} = \textbf{A}\textbf{Y} \sim {\N}\big( \textbf{0},{\varvec{\upSigma}}(\textbf{u})\big) $.

A second the difficulty is to find an appropriate decomposition of $ \varvec{\upSigma}(\textbf{u})=\textbf{A}(\textbf{u})\textbf{A}^T(\textbf{u}) $, in order to proceed later with the decoupling of $ \textbf{Z} $ and be able to work with independent variables $\textbf{Y}(\textbf{u})  =  \textbf{A}(\textbf{u})^{-1}\textbf{Z}(\textbf{u}) \sim \N(\textbf{0},\textbf{I}_p)$. One can suggest the use of eigen-decomposition $ \varvec{\upSigma} = \textbf{U}\textbf{D}\textbf{U}^T$ in order  $ \textbf{Z} = \textbf{U}{\textbf{D}}^{1/2}\textbf{Y}$, but used in automatized way may result in a model with spatial discontinuities, since the non-uniqueness of this decomposition. $ \varvec{\upSigma} $ can be uniquely decomposed as the product
of a positive-diagonal lower triangular matrix by Cholesky decomposition, being a suitable choice for our purposes: $ \varvec{\upSigma} = \textbf{L}\textbf{L}^T$.	

Once getting a continuous decomposition for $\varvec{\upSigma}(\textbf{u})$ and the independent variables, the overall process of estimation and simulation becomes straightforward, by working individually on the spatial behavior in each of the variables separately.

%%%%%%%%%%%%%%%%%%%%%%%%%%%%%%%%%%%%%%%%%%%%%%

\begin{figure}[htbp]
	\centering 
	\raggedright
	\includegraphics[height=0.2\textwidth,trim={0 0cm 0.0cm 0cm},clip]{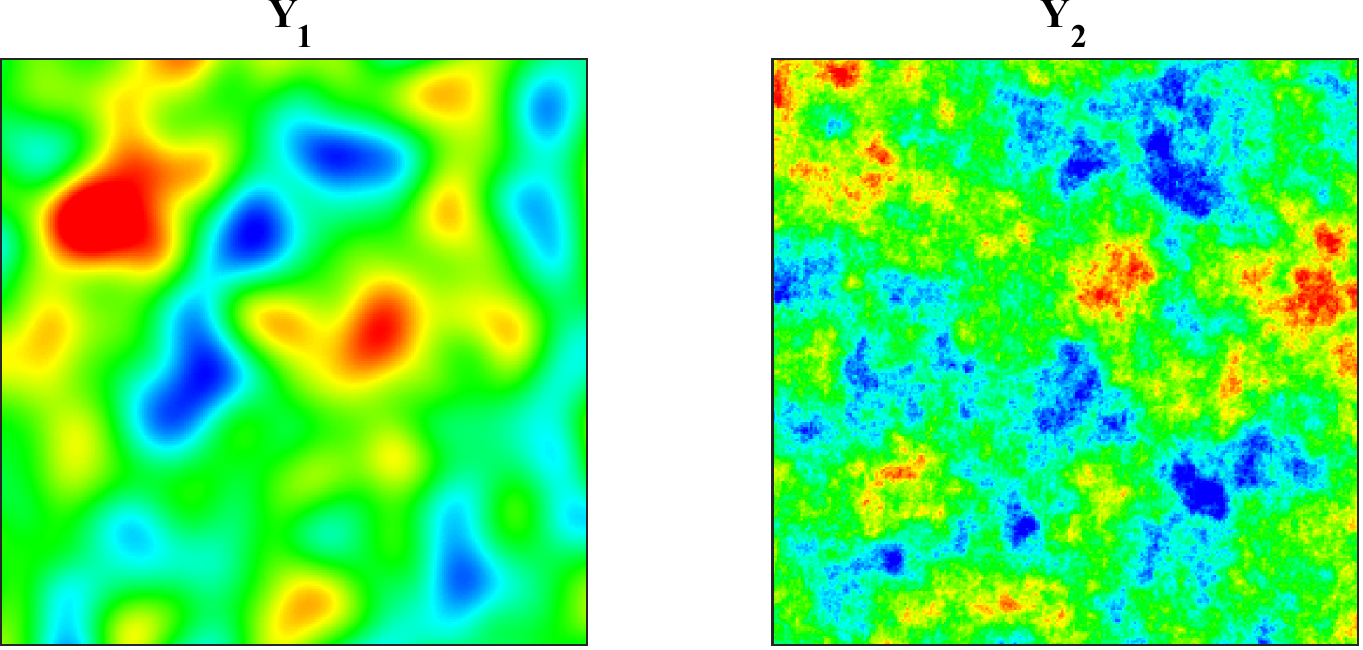}$ \quad $
	\includegraphics[height=0.19\textwidth,trim={0 0cm  0.0cm 0cm},clip]{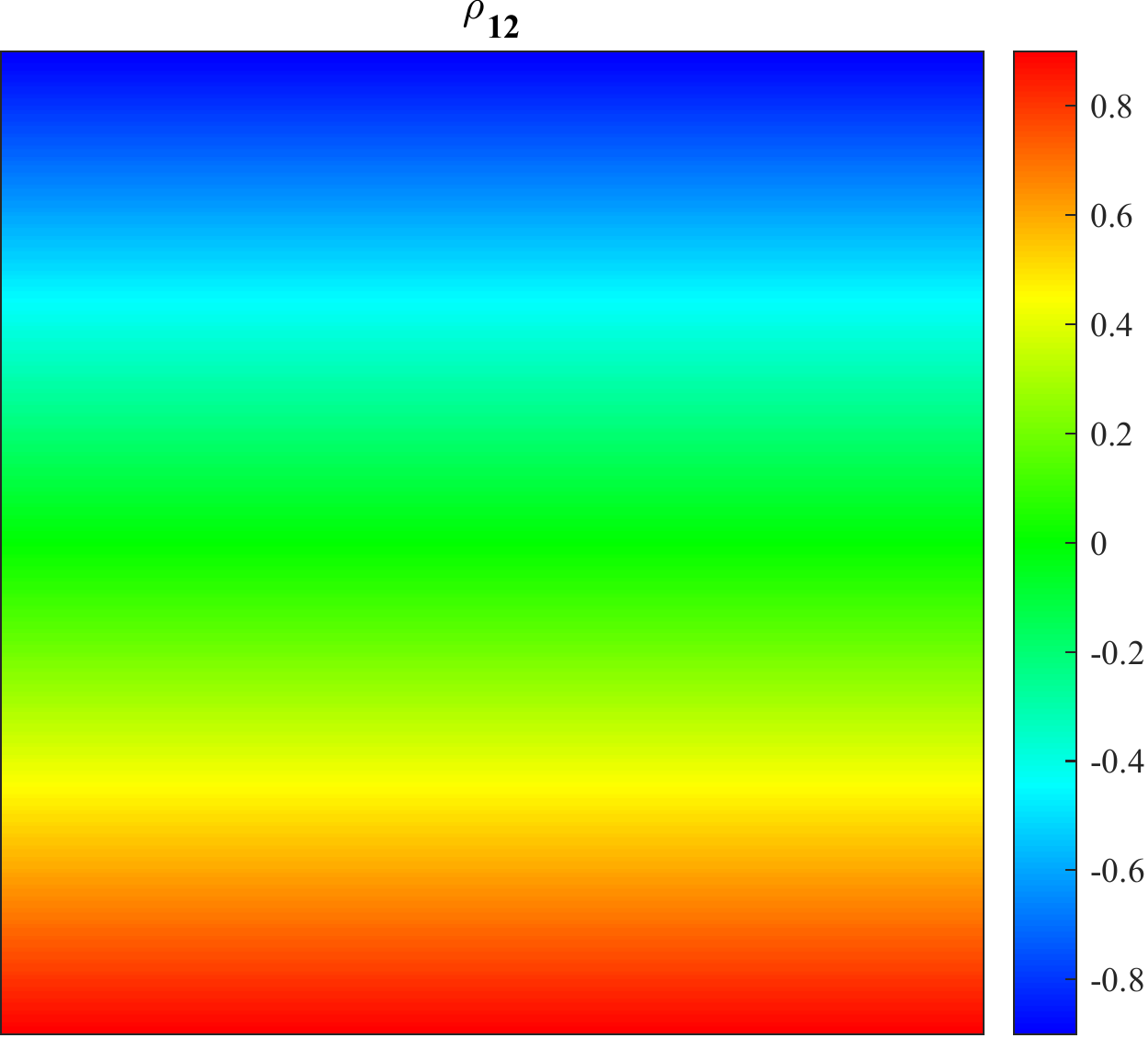} $ \quad 
	\begin{pmatrix}
		1 & \rho_{12}(\textbf{u})\\
		\rho_{12}(\textbf{u}) & 1\\
	\end{pmatrix} $ $\quad \,\,\,\,$\\
	\raggedright
	\includegraphics[,width=0.42\textwidth]{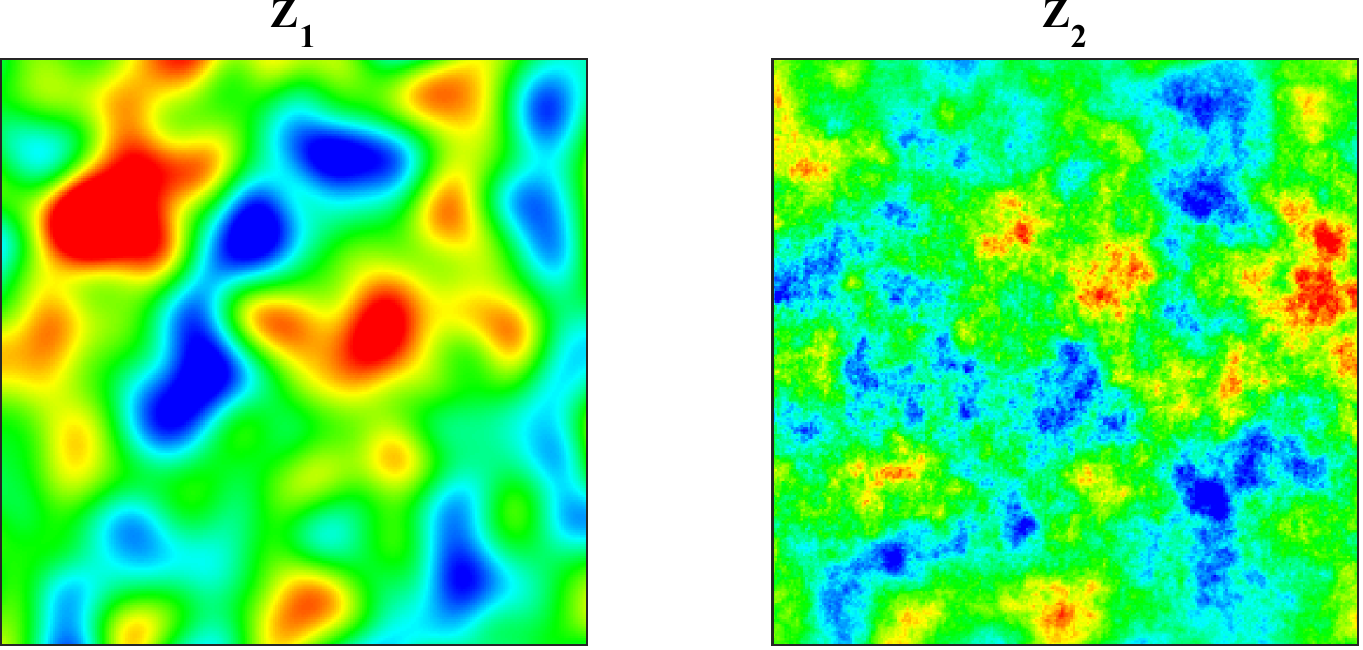}\\
	\includegraphics[,width=0.42\textwidth]{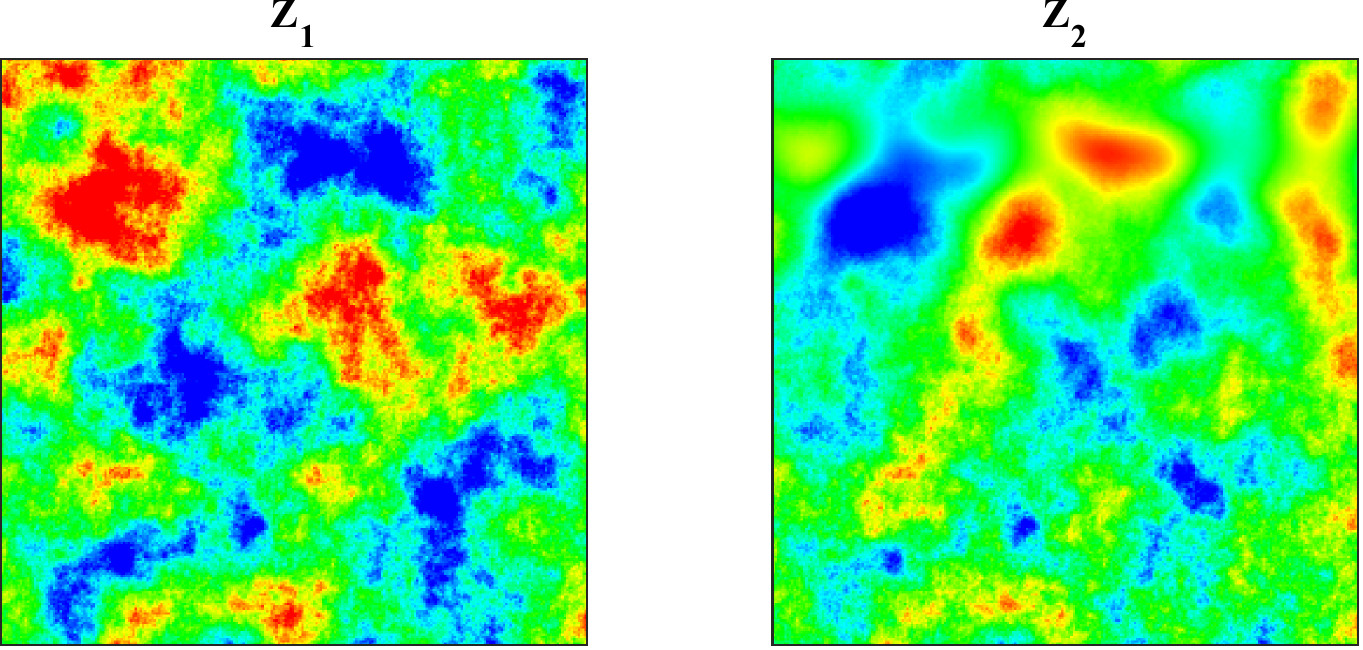}
	\caption{Effect of the rotation and mixing of two independet Gaussian RF $\textbf{Y} = [ Y_1(\textbf{u}) \quad Y_2(\textbf{u})]^T \sim \N(\textbf{0},\textbf{I}_2) $ with different spatial continuity (top) to get $ \textbf{Z} = \textbf{L}\textbf{R}\textbf{Y}$ for two different options of $ \textbf{R} $ (middle and bottom), and taking $ \textbf{L} $ the Cholesky matrix for a given correlation map shown on top.}
	\label{fig:rotat}
\end{figure}

One last difficulty in overcoming comes  from the fact that, once Cholesky is applied, one notices that the following transformation also works well:	$ \textbf{Z} = \textbf{L}\textbf{R}\textbf{Y}$, with $ \textbf{R} $ a rotation matrix, as any decomposition of the form $ \varvec{\upSigma} = \textbf{L}\textbf{R}\textbf{R}^T\textbf{L}^T$ is valid. This is a bit problematic since the model acquires an extra free parameter, which is a source for ill-definition for our model if different spatial models are involved in the $ Y_i(\mathbf{u}) $ variables (Fig. \ref{fig:rotat}). If there is a way for finding a suitable $ \textbf{R} $ and fixing this parameter, that is a topic of further research. In order to further simplify these issues and the methodology overall at this point, we take $ Y_i(\mathbf{u}) $ following the same variogram model for all $ i \in I $.

Now we proceed to deal with the problem of interpolating the different known correlations matrices in the space.

\subsubsection{Interpolation of the Correlation Matrices}

We present a fixed point and a gradient descent algorithm which seeks to minimize the mean-squared distances of $ \textrm{Sym}^+(p) $ and $ \textrm{Corr}(p) $-valued observations, respectively, with respect to the affine-invariant distance. The general
process for the optimization procedure for the $ \textrm{Corr}(p) $ is proposed by \cite{david2019riemannian}, and takes the following steps:

\begin{enumerate}
	\item At the current iterate $ \textbf{C}_t \in \textrm{Corr}(p) $ find all appropriate distances utilizing the fiber structure of $\textrm{Sym}^+/\textmd{Diag}^{+} $.
	\item  Interpret $ \textbf{C}_t \in \textrm{Sym}^+(p) $ and perform the update to a point $\textbf{P}_{t+1}\in \textrm{Sym}^+(p) $.
	\item Obtain the next iterate in the algorithm by projecting back to $ \textrm{Corr}(p) $, that is $ \textbf{C}_t = \pi(\textbf{P}_{t+1}) $.
\end{enumerate}

We begin by summarizing the optimization methods on $ \textrm{Sym}^+(p) $ and $ \textrm{Corr}(p) $.

\subsubsubsection{Optimizing on $ \textrm{Sym}^+(p) $}

Given the observations  $ \textbf{P}_1,\dots,\textbf{P}_k \in \textrm{Sym}^+(p) $, one could consider the \textit{arithmetic mean} of the $ n $ labeled covariance matrices $ \{\textbf{P}_i\}^{k}_{i=1} $: \[ \widehat{\varvec{\upSigma}} = \frac{1}{k}\sum_{i=1}^{k}\textbf{P}_i \]which do not account for any intrinsic geometric property of $ \textrm{Sym}^+(p) $.

We consider, instead, to use the \textit{\textbf{geometric}} or \textbf{\textit{Fr\'echet mean}}, introduced in the $ \textrm{Sym}^+(p) $ context by \cite{moakher2005differential}. Such a matrix is defined as follows:
\begin{equation}\widehat{\varvec{\upSigma}} = \textmd{arg} \inf_{\substack{\varvec{\upSigma}}} \sum_{i=1}^k d^2_{\textrm{Sym}^+}(\textbf{P}_i,\varvec{\upSigma}).\label{minim}
\end{equation}
Recall that the Riemannian distance between two SPD matrices is defined as:
\begin{eqnarray} d_{\textrm{Sym}^+}^2(\textbf{P}_i, \varvec{\upSigma})= \norm{\textmd{log}_{\textbf{P}_i}(\varvec{\upSigma})}^2 &=& \norm{\text{Log}(\textbf{P}^{1/2}_i\varvec{\upSigma}\textbf{P}^{1/2}_i)}^2 \nonumber \\
&=& \text{tr}\big(\text{Log}^2(\textbf{S}^{1/2}_i\varvec{\upSigma}\textbf{S}^{1/2}_i)\big)\nonumber
\end{eqnarray}
and, therefore, minimizing Eq. \ref{minim} seems to be impossible to solve in closed form, according to \cite{moakher2006averaging}. The same author describe a fixed-point algorithm to numerically solve the geometric mean of a set of symmetric positive-definite matrices. Other methods such as Newton’s method on Riemannian manifolds \citep{david2019riemannian} could also be used for the numerical
computation of the geometric mean. However, the fixed-point
algorithm described below is simple to implement, does not require a
sophisticated machinery, and converges rapidly.

The geometric mean $ \widehat{\varvec{\upSigma}} $ can be computed efficiently by an iterative procedure consisting in: projecting the covariance matrices in the tangent space, estimating the arithmetic mean in the tangent space and projecting the arithmetic mean back in the manifold. Then iterate the three above steps until convergence.

If we want to account for the spatial configuration of the data, we need to consider the use of the \textbf{\textit{weighted Fr\'echet mean}}: \[ \widehat{\varvec{\upSigma}} = \textmd{arg} \inf_{\substack{\varvec{\upSigma}}} \sum_{i=1}^n \lambda_i d^2_{\textrm{Sym}^+}(\textbf{S}_i,\varvec{\upSigma}), \quad \sum_{i=1}^n \lambda_i = 1,\]
with $ \lambda_i $ the weights obtained from the kriging interpolation. The algorithm in this case is given by slightly modifying the one taken from \cite{moakher2006averaging}:

\begin{algorithm}
	\caption{Weighted mean of $ k $ SPD matrices }\label{alg:cap2}
	\begin{algorithmic}[1]
		\Require a set of $ k $ SPD matrices  $ \textbf{P}_1,\dots,\textbf{P}_k \in \textrm{Sym}^+(p) $  and $ \epsilon> 0$.
		\State Initialize $\widehat{\varvec{\upSigma}}^{(1)} = \sum_{i=1}^{k}\lambda_i\textbf{P}_i$
		\Repeat
		\State ${\varvec{\bar P}} = \sum_{i=1}^{k}\lambda_i\textmd{log}_{\widehat{\varvec{\upSigma}}^{(t)}}(\textbf{P}_i)$ \Comment{Weighted mean in the tangent space}
		\State $\widehat{\varvec{\upSigma}}^{(t+1)}=\textmd{exp}_{\widehat{\varvec{\upSigma}}^{(t)}}({\varvec{\bar P}})$  
		\Until{$ \norm{{\varvec{\bar P}}} < \epsilon$}\\
		\Return $\widehat{\varvec{\upSigma}}^{(t+1)}$	
	\end{algorithmic}
\end{algorithm}

\subsubsubsection{Optimizing Along Fibers}

In the same fashion as previously, given the observations  $ \textbf{C}_1,\dots,\textbf{C}_k \in \textrm{Corr}(p) $, we are interested in finding

\begin{equation}\widehat{\textbf{C}} = \textmd{arg} \inf_{\substack{\textbf{C}}} \sum_{i=1}^k \lambda_id^2_{\textrm{Corr}}(\textbf{C}_i,\textbf{C}), \quad \sum_{i=1}^n \lambda_i = 1.\label{minim2}
\end{equation}

Recall again that the distance between $ \textbf{C}_i, \textbf{C} \in \textmd{Corr}(p) $ is given
by

\begin{eqnarray}
	d^2_{\textmd{Corr}}(\textbf{C}_i,\textbf{C})&= &\inf_{\substack{\textbf{D} \in \textmd{Diag}^{+}(p)}}d^2_{\textrm{Sym}^+}(\textbf{C}_i,\textbf{DC}\textbf{D}) \nonumber\\
	&= &\inf_{\substack{\textbf{D} \in \textmd{Diag}^{+}(p)}}\textmd{tr}\big[\textmd{Log}^2\big(\textbf{C}_i^{-1/2}\textbf{DC}\textbf{D}\textbf{C}_i^{-1/2}\big)\big], \nonumber
\end{eqnarray}
where we note that one can fix $ \textbf{C} $ and then optimize over the fiber of $ \textbf{C}_i $ as well, by symmetry. The minimization of the distance between an iterate $ \textbf{C}_t $
of the algorithm to all of the observations $\textbf{C}_1,\dots,\textbf{C}_k $ is preferred. Hence, the algorithm is arranged to always keep the iterate fixed and then optimizing along the fibers of the given observations. In this way, it is guarantee that the itererated point is updated appropriately. In other case, one would end up with different optimal points, not yielding a consistent base point. In finding the optimal point, it is employed a gradient descent method on the set $ \textmd{Diag}^+(p) $ with respect to the objective function
\[ g_i(\textbf{D})= d^2_{\textrm{Sym}^+}(\textbf{C},\textbf{DC}_i\textbf{D})\]

The gradient descent algorithm in order to find the optimal $ \textbf{D} $ in the above expression is proposed by \cite{david2019riemannian}. The algorithm's derivation is long and tedious, and we refer to the mentioned author for further details. One ends up, however, with a brief two-steps iterative algorithm, by using a stepsize $ \delta > 0  $, initializing $ \textbf{D}_0 = \textbf{I}_p $ and following iterative steps
\begin{eqnarray}
	\varvec\Delta_t & = &\textbf{I} \circ 2\textmd{Sym}[\textbf{D}_t\textmd{Log}(\textbf{C}_i\textbf{D}_t\textbf{C}^{-1}\textbf{D}_t)],\nonumber\\
	\textbf{D}_{t+1} & = & \textbf{D}_t\textmd{Exp}(-\delta\textbf{D}^{-1}_t\varvec\Delta_t),\nonumber
\end{eqnarray}
with $ \textmd{Sym}(\textbf{A})=\frac{1}{2}(\textbf{A}+\textbf{A}^{T}) $, until a desired stopping criterion is reached. Once we find an optimal element $\textbf{D}^* \in \textmd{Diag}^{+}(p)$ as a result of minimizing $ g_i(\textbf{D}) $,  we define as $ \widetilde{\textbf{C}}_i $ this element over the fiber $ \pi^{-1}(\textbf{C}_i) $ which minimizes the $ {\textrm{Sym}^+} $-distance between $ \textbf{C} $ and $ \textbf{C}_i $,  ${\widetilde{\textbf{C}}}_i = \textbf{D}^*\textbf{C}_i\textbf{D}^*$.

We summarize the proposed algorithm which finds the Fr\'echet mean on $ \textmd{Corr}(p) $:

\begin{algorithm}
	\caption{Weighted mean of $ k $ Correlation matrices }\label{alg:cap3}
	\begin{algorithmic}[1]
		\Require a set of $ k $ Correlation matrices  $ \textbf{C}_1,\dots,\textbf{C}_k \in \textrm{Corr}(p) $ and $ \epsilon> 0$, initial point $ \textbf{C}_0= \sum_{i=1}^{k}\lambda_i\textbf{C}_i$, stepsize $ \delta>0 $.
		\State $ t=0 $
		\While {Stopping criterion not met}
		\For{$ i= 1,\dots,k$}
		\Require Initial point $ \textbf{D}_0 $
		\State $ n= 1$
		\While {Stopping criterion not met}
		\State $ \varvec\Delta_t =\textbf{I} \circ 2\textmd{Sym}[\textbf{D}_n\textmd{Log}(\textbf{C}_i\textbf{D}_n\textbf{C}^{-1}_t\textbf{D}_n)] $
		\State $ \textbf{D}_{n+1} = \textbf{D}_{n}\textmd{Exp}(-\delta\textbf{D}^{-1}_n\varvec\Delta_n) $
		\State $ n = n+1 $
		\EndWhile
		\State ${\widetilde{\textbf{C}}}_i = \textbf{D}_{n_{max}}\textbf{C}_i\textbf{D}_{n_{max}}$
		\EndFor
		\State ${\varvec{\bar P}}_t = \sum_{i=1}^{k}\lambda_i\textmd{log}_{\textbf{C}_t}({\widetilde{\textbf{C}}}_i)$ \Comment{Mean in the tangent space of $ \textmd{Sym}^{+}(p) $}
		\State ${\varvec{\upSigma}}_{t+1}=\textmd{exp}_{{\textbf{C}_t}}({\varvec{\bar P}})$
		\State $ \textbf{C}_{t+1} =\pi({\varvec{\upSigma}}_{t+1})= (\textbf{I}_p \circ \varvec{\upSigma}_{t+1})^{-1/2}\varvec{\upSigma}_{t+1} (\textbf{I}_p \circ \varvec{\upSigma}_{t+1})^{-1/2} $
		\Comment{Project back to $ \textmd{Corr}(p) $}
		\State $ t=t+1 $
		\EndWhile\\
		\Return $\textbf{C}_{t}$	
	\end{algorithmic}
\end{algorithm}

\subsubsection{Methodology}

Now that we have gone throughout the steps for interpolating the correlation matrices, we summarized the proposed methodology for extending the LMC, consisting of the following steps with both the first and the last step being optional and suggested when the data is compositional:

\begin{enumerate}[label=\arabic*)]
	\item (Perform log-ratio transformation on data, if constrains conditions are present).
	\item At each location $ \textbf{u}_\alpha $ with observation, find the nearest $ l $ samples.
	\item Perform Gaussian transformation individually for each variable $\phi_i^{-1}[\tilde Z_i(\textbf{u}_\alpha)]={Z}_i(\textbf{u}_\alpha)$ locally using the nearest $ l $ samples.
	\item Compute the correlation matrix $ \varvec{\upSigma}(\textbf{u}_\alpha) $ of the vector $ {\textbf{Z}}=[{Z}_1(\textbf{u}_\alpha),\dots,{Z}_p(\textbf{u}_\alpha)] $.
	\begin{enumerate}[label=\alph*)]
		\item Cholesky decomposition of $ \varvec{\upSigma}(\textbf{u}_\alpha)= \textbf{L}\textbf{L}^T $ and apply $ \textbf{L}^{-1}{\textbf{Z}}={\textbf{Y}} $ for decorrelation of Gaussian variables.
		\item Interpolation of $ \varvec{\upSigma}(\textbf{u}_\alpha)$ on the domain $ D $ using weighted Fr\'echet mean and your favorite set weights $ \lambda_i $. Kriging weights given by the variogram modeling of $ {\textbf{Y}} $ works appropriately.
	\end{enumerate}
	\item Variogram modeling and simulation of $ {{Y}_i} $, assuming the same model $ \forall i \in I $.
	\item At each unsampled location $ \textbf{u} $, take the estimated correlation matrix $ \widehat{\varvec{\upSigma}}(\textbf{u}) $, perform cholesky decomposition, and recover $ {\textbf{Z}}(\textbf{u})=\widehat{\textbf{L}}(\textbf{u}){\textbf{Y}}(\textbf{u}) $.
	\item At the unsampled location $ \textbf{u} $, find the nearest $ l $ samples, perform Gaussian transformation individually for each variable $\widehat{\phi}_i^{-1}[\tilde Z_i(\textbf{u})]={Z}_i(\textbf{u})$, and recover the value $ \tilde z_i(\textbf{u})=\widehat{\phi}_i[z_i(\textbf{u})] $.
	\item (Perform  log-ratio back transformation on data).
\end{enumerate}

\subsection{Geological Domaining}

Now we move to a second novel application of the concepts presented on previous chapters. We are considering the classical problem in geostatistics of clustering data which carry continuous information in space, $z(\textbf{u}$) (such as a grades), where $\textbf{u}$ is the vector in the three-dimensional space ($\textbf{u} \in \R^3$). However, $z$ has only been sampled in a discrete set of points $\{\textbf{u}_{\alpha},\alpha \in N=1,\dots,n\}$. From these measurements, we have some intuition that there is an unknown finite collection $ A_1, A_2,\dots$ , $ A_k $ of disjoint sets of  $N $, with $ \cup_{i=1}^{k}A_i=N $, on which the measurement $z(\textbf{u}_{\alpha}),\alpha \in A_l$ and $z(\textbf{u}_{\beta}),\beta \in A_m$, for all $ l\neq m $, have low relationship (or not at all) between them (for example, because they have a different genesis) and, therefore, they should be clustered on different categories (typical examples are lithofacies types). We want to find the collection $ A_1, A_2,\dots, A_k $.

Methodologies able to deal with this problem have several significant applications. One of the most important is the definition of stationary spatial domains, where the assumption of a fairly constant mean within a given spatial domain is critical for some aspects of resource estimation. There is no resource estimation done without the definition of stationary units. This process is known as the definition of \textit{geological units} in geoscientific terms. Most of the time, geological domaining is done based on non-continuous attributes (lithology and alteration of the rock) which is related or explain somehow the values of continuous data. Sometimes, however, the categorical information is not enough to define the units by itself.

The methodology presented next implements $ K$-means algorithm on $ \textmd{Corr}(p) $ and is an alternative to include the spatial information on continuous data, and should help the geo-modeler to decide boundaries for geological units, in cases of fuzzy or contradictory categorical data.

\subsubsection{Methodology} 

Following an idea proposed in \cite{you2021re} for the SPD case, we implemented $K$-means algorithm modified to our
context. $K$-means algorithm \citep{macqueen1967some} is one of famous clustering algorithms for data analysis. As pointed out in \cite{goh2008clustering}, the method is easily extensible to non-Euclidean data as it solely depends on the distance measure in determining class memberships.

\begin{enumerate}[label=\arabic*)]
	\item  randomly choose $ K $ correlations matrices as cluster means, $ \varvec\mu^{(1)}_1,\dots,\varvec\mu^{(1)}_K $, where the upper index refers to the number of iteration.
	\item repeat following steps until convergence:
	\begin{enumerate}[label=\alph*)]
		\item assign each observation to the cluster by smallest distances to
		cluster centers,\[ S^{(t)}_i =\{\varvec\upSigma_\alpha : d(\varvec\upSigma_\alpha,\varvec\mu^{(t)}_i)\}\leq d(\varvec\upSigma_\alpha,\varvec\mu^{(t)}_j) \textmd{ for all } 1\leq j\leq K\}\]
		and when it comes to a situation where an observation can belong
		to one of multiple clusters, assign the cluster randomly.
		\item update cluster centroids by Fréchet means of each class,\[\varvec\mu^{(t+1)}_i = {\underset {\textbf{P} \in \textmd{Corr}(p)}{\operatorname {arg\,min} }}\,\sum_{j \in S^{(t)}_i} d^2(\textbf{P},\varvec\upSigma_\alpha)  \textmd{ for } i=1,\dots,K \]
	\end{enumerate}
\end{enumerate}

\begin{figure}[h!]
	\centering
	\includegraphics[width=0.325\textwidth,trim={0 0cm 0.0cm 0cm},clip]{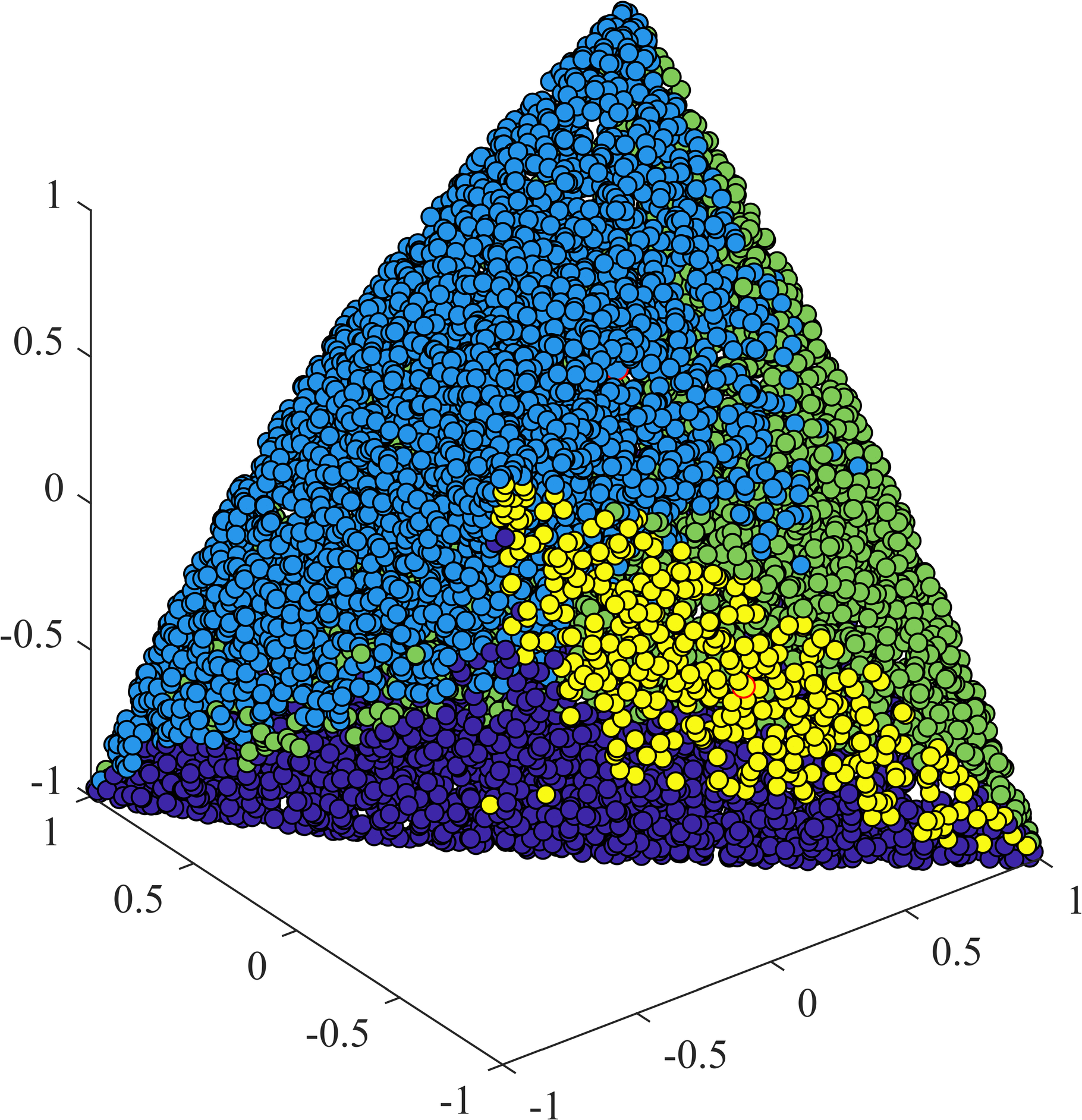}
	\includegraphics[width=0.325\textwidth,trim={0 0cm 0.0cm 0cm},clip]{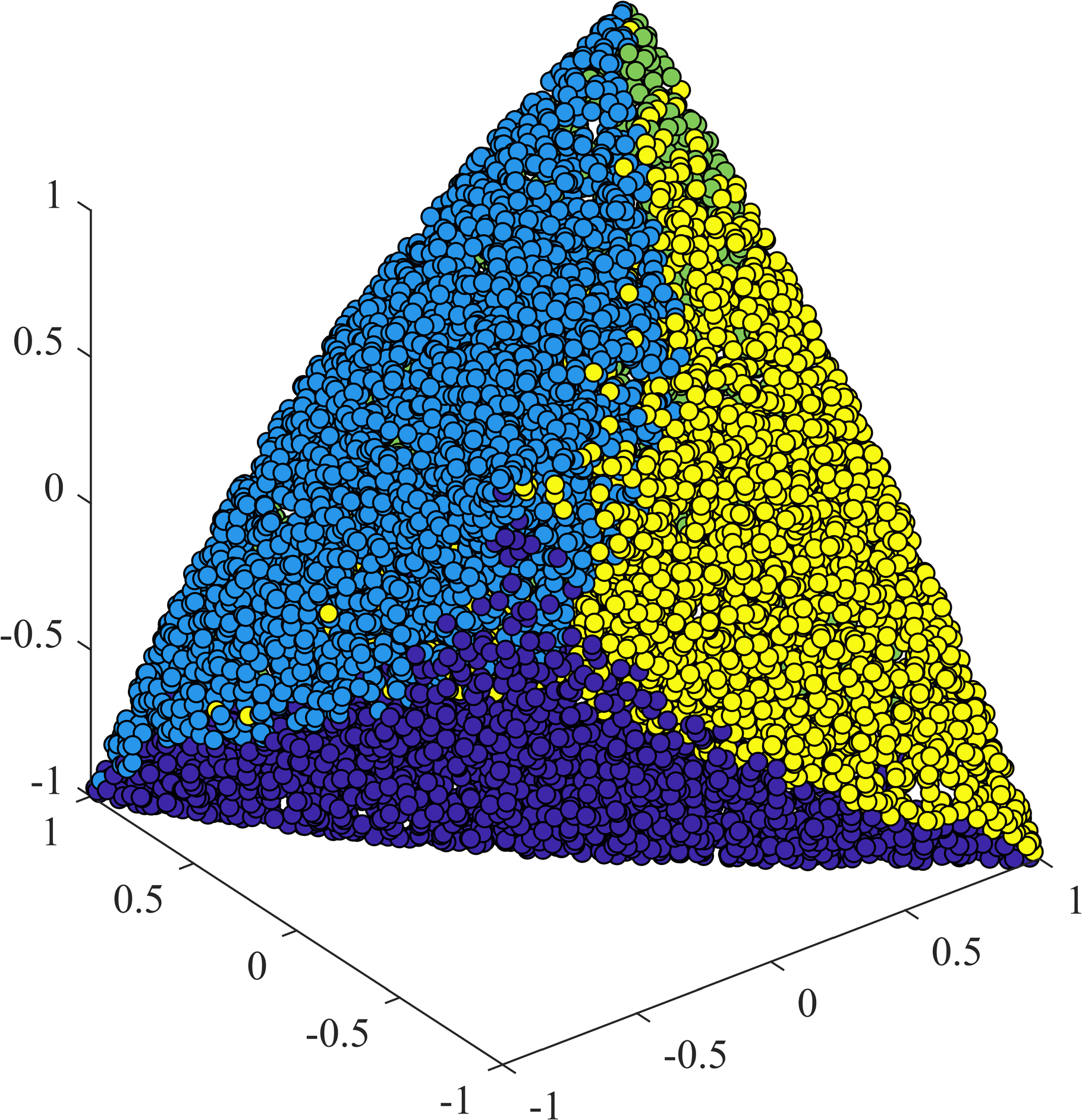}
	\includegraphics[width=0.325\textwidth,trim={0 0cm 0.0cm 0cm},clip]{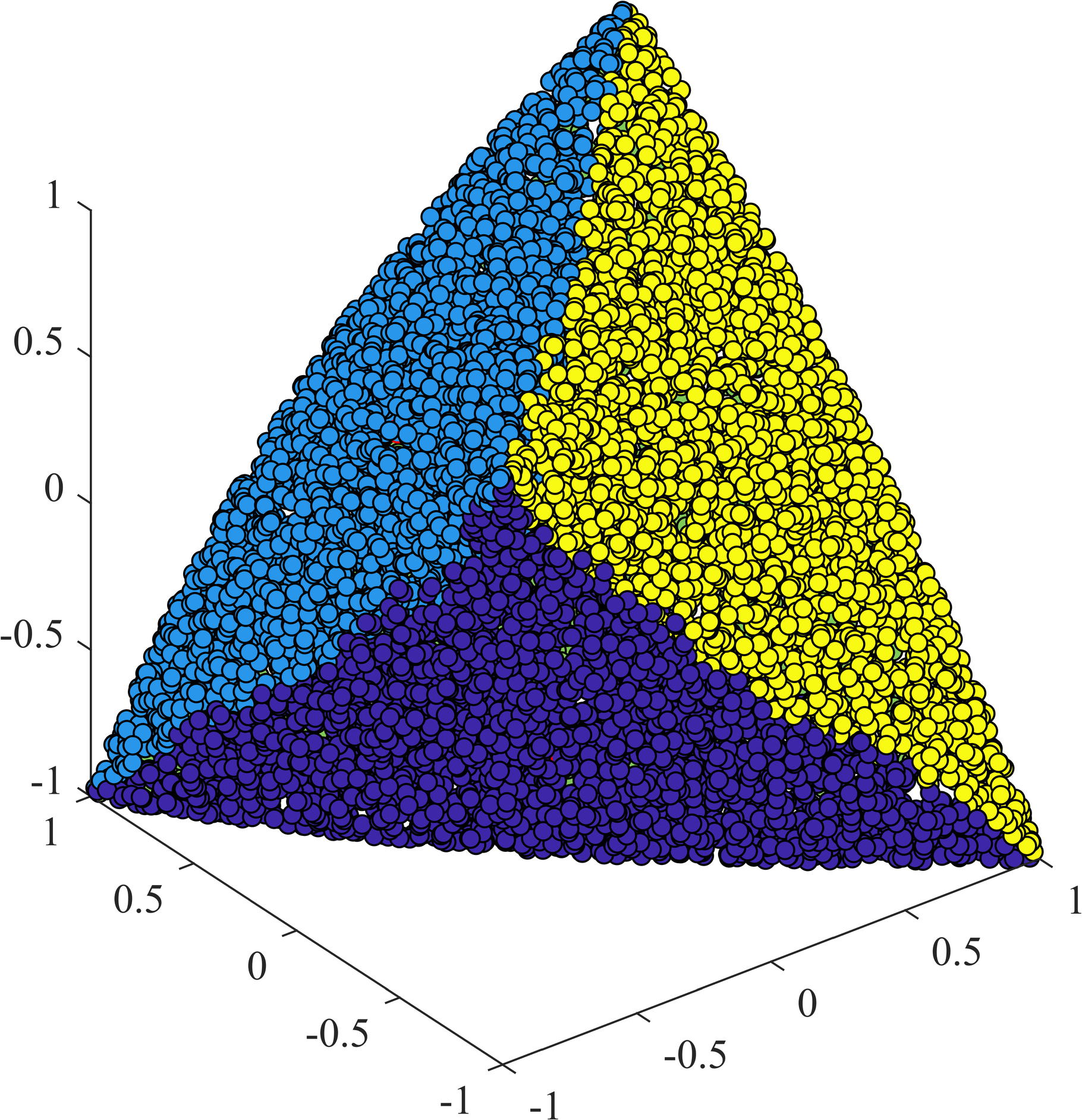}
	\caption{Clustering on random samples of the $\textrm{Corr}(3) $ manifold, and using $ K$-means algorithm with $ K=4 $. Iterations 1, 2 and 5 are displayed from the top, respectively.}
	\label{corrmanclus}
\end{figure}

\begin{figure}[h!]
	\centering
	\includegraphics[width=0.325\textwidth,trim={0 0cm 0.0cm 0cm},clip]{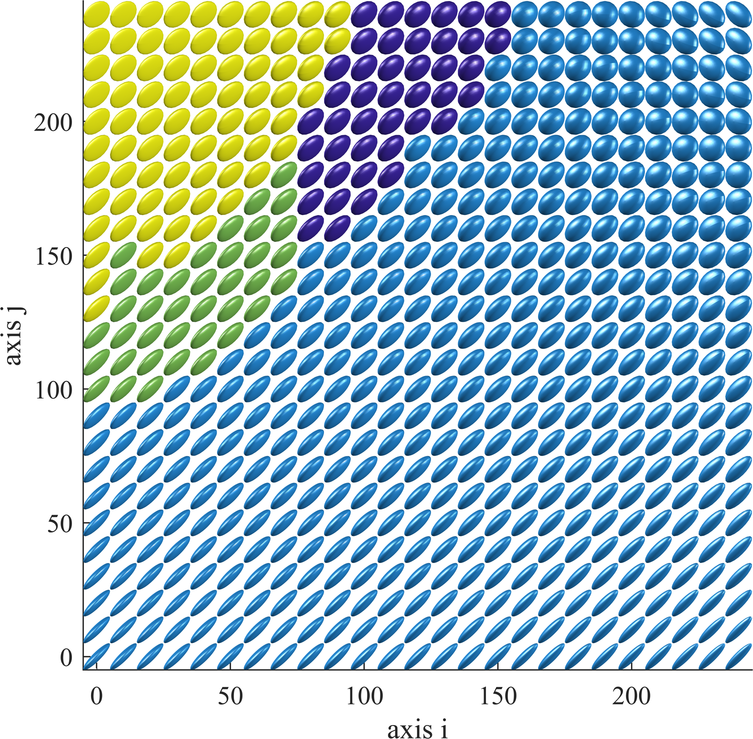}
	\includegraphics[width=0.325\textwidth,trim={0 0cm 0.0cm 0cm},clip]{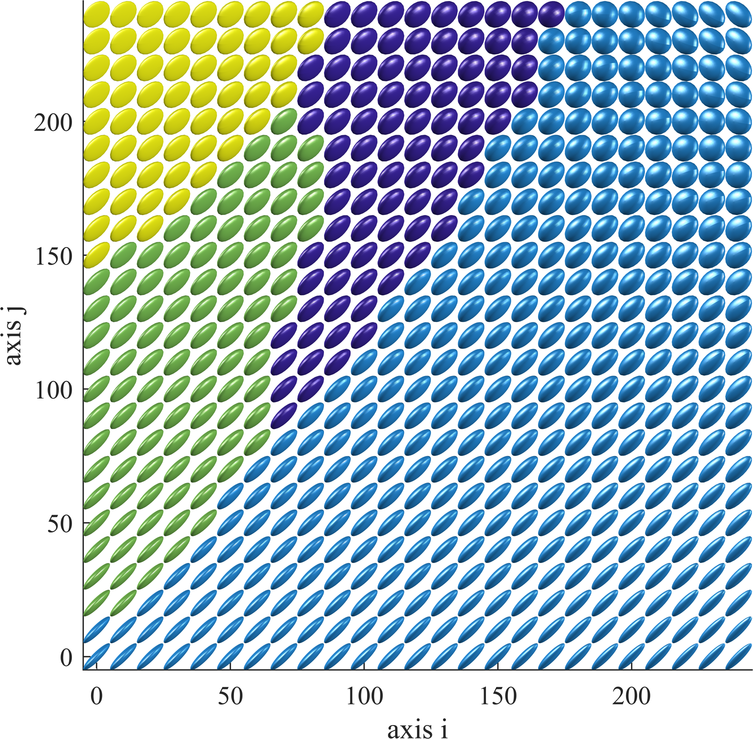}
	\includegraphics[width=0.325\textwidth,trim={0 0cm 0.0cm 0cm},clip]{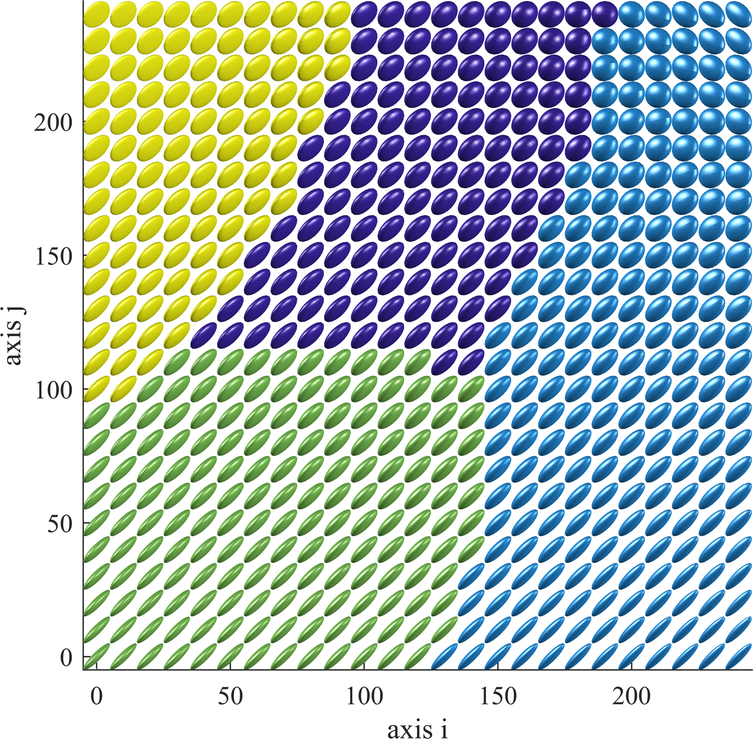}
	\caption{Clustering on the $\textrm{Corr}(3) $ manifold using $ K$-means algorithm with $ k=4 $, on a distribution of interpolated ellipses on the space. Iterations 1, 2 and 11 are displayed from the top, respectively.}
	\label{corrmanclus2}
\end{figure}

We have tested the algorithm on $ \textmd{Corr}(3) $, that is, on the elliptope (Fig. \ref{corrmanclus}), and later, on a set of interpolated correlation matrices on the space (Fig. \ref{corrmanclus2}), both with good results in terms of the continuity of the clusters.

\section{Case Study}
\label{Case Study}

\subsection{The Data}

In order to show the capabilities of the proposed techniques described previously, a data set obtained from a blast hole campaign pertaining to a Nickel-Laterite deposit is considered and six cross-correlated variables  isotopically assayed at each sample point: Fe, Ni, MgO, SiO$_2$, Al$_2$O$_3$, and Cr. Isotopic sampling ensures that all the variables are available through all the sample locations \citep{wackernagel2013multivariate}. The case study includes 9990 samples available on the data set with a very dense sampling pattern. The name and location of data set cannot be disclosed because of confidentiality reasons.

\begin{figure}[h!]
	\centering
	\includegraphics[width=0.45\textwidth,trim={0 0cm 0.0cm 0cm},clip]{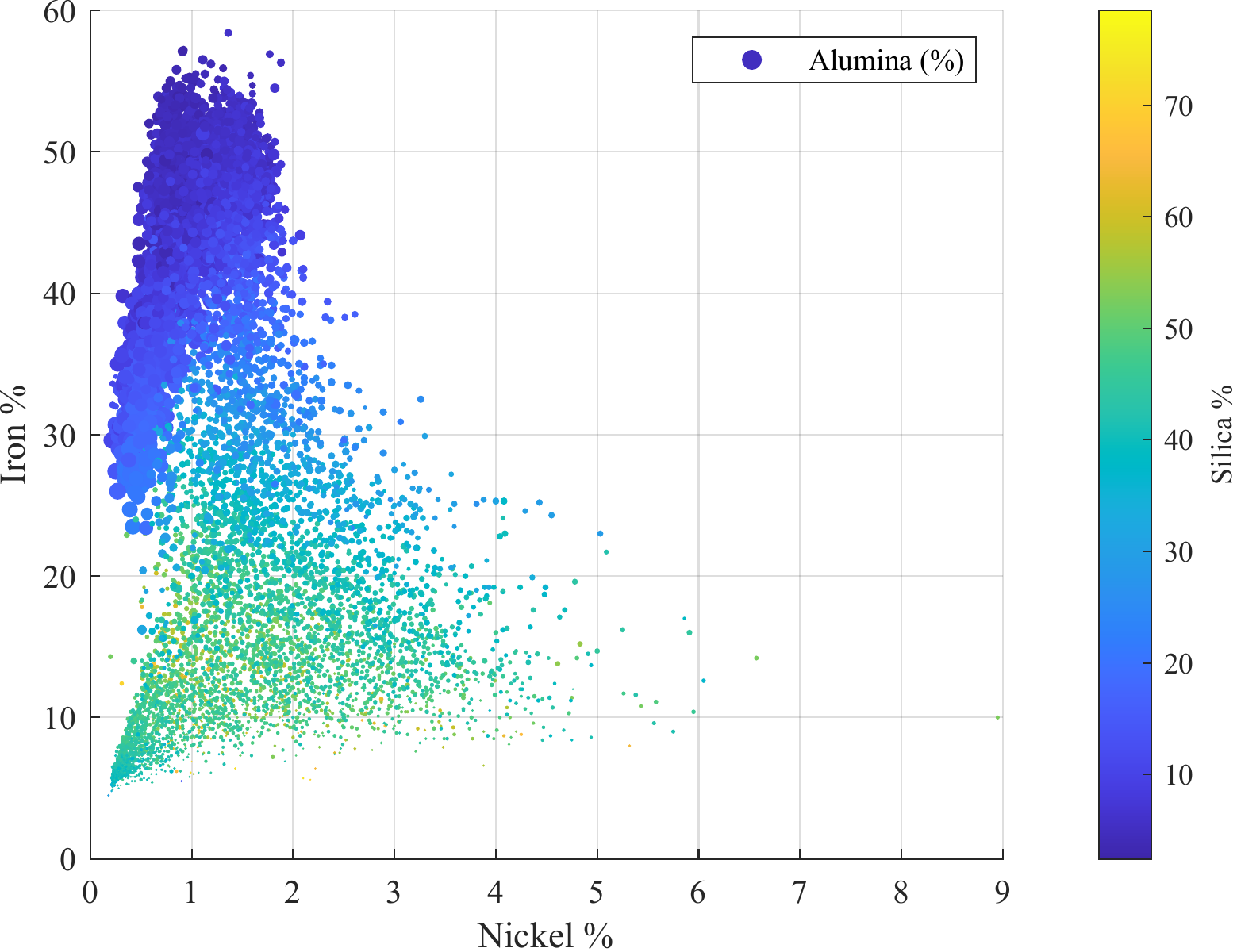}
	\includegraphics[width=0.45\textwidth,trim={0 0cm 0.0cm 0cm},clip]{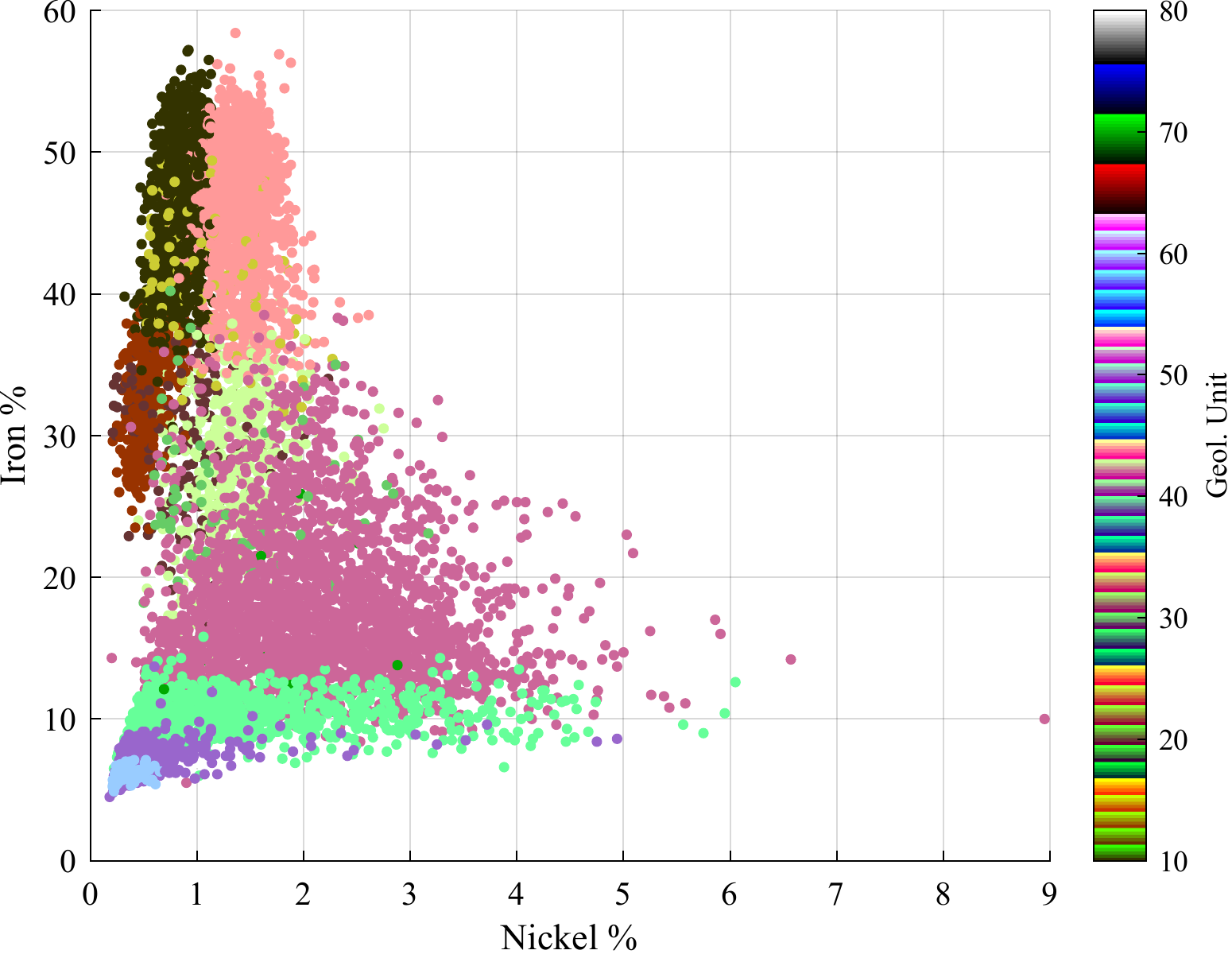}
	\caption{Display of multivariate features on sampling data of Nickel-Laterite Deposit. 4 out of 6 variables can be seen on the scatter plots, by adding color and a variable diameter to the bullet, proportional to the amount of alumina (top). Geological codes provided are display as well (bottom).}
	\label{loc2}
\end{figure}

\begin{figure}[h!]
	\centering
	\includegraphics[width=0.49\textwidth,trim={0 0cm 0.0cm 0cm},clip]{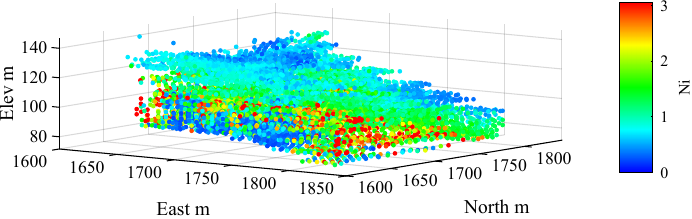}
	\includegraphics[width=0.49\textwidth,trim={0 0cm 0.0cm 0cm},clip]{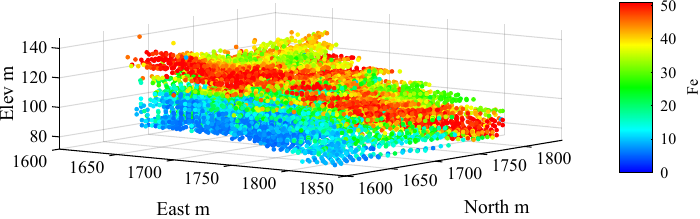}
	\includegraphics[width=0.49\textwidth,trim={0 0cm 0.0cm 0cm},clip]{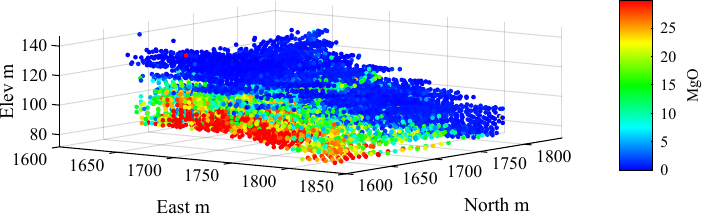}
	\includegraphics[width=0.49\textwidth,trim={0 0cm 0.0cm 0cm},clip]{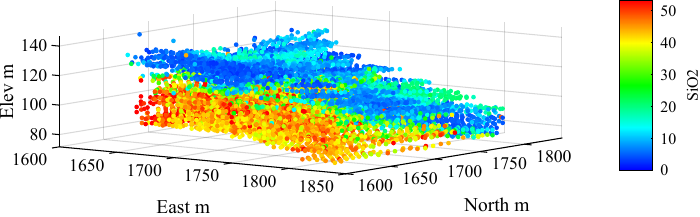}
	\includegraphics[width=0.49\textwidth,trim={0 0cm 0.0cm 0cm},clip]{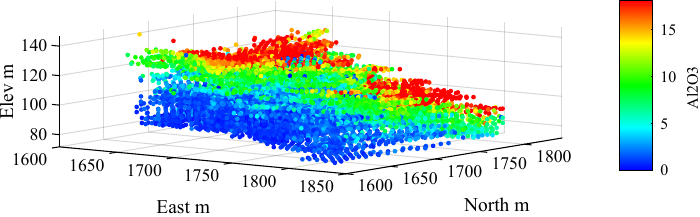}
	\includegraphics[width=0.49\textwidth,trim={0 0cm 0.0cm 0cm},clip]{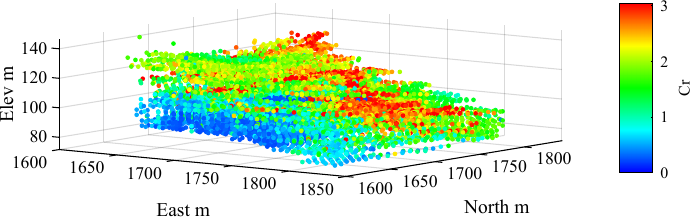}
	\caption{Isometric view showing the sampling grade information.}
	\label{loc14}
\end{figure}

The primary inspection of multivariate relations (scatter-plot shown on Fig. \ref{loc2}) exposed many aspects of complexity such as non-linearity and heteroscedasticity. A map of the samples for each variable is presented in Fig. \ref{loc14}. In order to show the predictability of the proposed methodology, 500 random samples are selected and taken away for testing purposes.

%\begin{figure}[h!]
%	\centering
%	\includegraphics[width=0.39\textwidth,trim={0 0cm 0.0cm 0cm},clip]{./fotos/hist1.pdf-1.png}
%	\includegraphics[width=0.39\textwidth,trim={0 0cm 0.0cm 0cm},clip]{./fotos/hist2.pdf-1.png}
%	\includegraphics[width=0.39\textwidth,trim={0 0cm 0.0cm 0cm},clip]{./fotos/hist3.pdf-1.png}
%	\includegraphics[width=0.39\textwidth,trim={0 0cm 0.0cm 0cm},clip]{./fotos/hist4.pdf-1.png}
%	\includegraphics[width=0.39\textwidth,trim={0 0cm 0.0cm 0cm},clip]{./fotos/hist5.pdf-1.png}
%	\includegraphics[width=0.39\textwidth,trim={0 0cm 0.0cm 0cm},clip]{./fotos/hist6.pdf-1.png}
%	\caption{Histogram for sampling data.}
%	\label{loc1}
%\end{figure}

\subsection{Variography}

As described in proposed methodology, we begin by applying additive log-ratio transformation on the data,  taken with respect to the Rest variable (Rest$ = 100$ \% $ - $ Ni\% $-\dots-$ Cr\%), as this extra variable may gives us further information in the prediction. Gaussian transformation is applied at each sample location, by selecting a neighborhood of the closest 800 samples. This parameter was calibrated several times, showing that working with less data reduces the capabilities for  reproduction of the multivariate behavior shown on the data drastically, as the correlation matrix gets distorted with a low amount of data.

Once the data is gaussianized and de-correlated after obtaining the correlation matrix, the experimental direct and cross omni-directional variograms are calculated. Variogram analysis in different directions was not considered as the amount of data in the vertical direction is much less than horizontally. This last aspect, however, is included later in the radius of search for estimation.

\begin{figure*}[h!]
	\centering
	\includegraphics[width=0.75\textwidth,trim={0 0cm 0.0cm 0cm},clip]{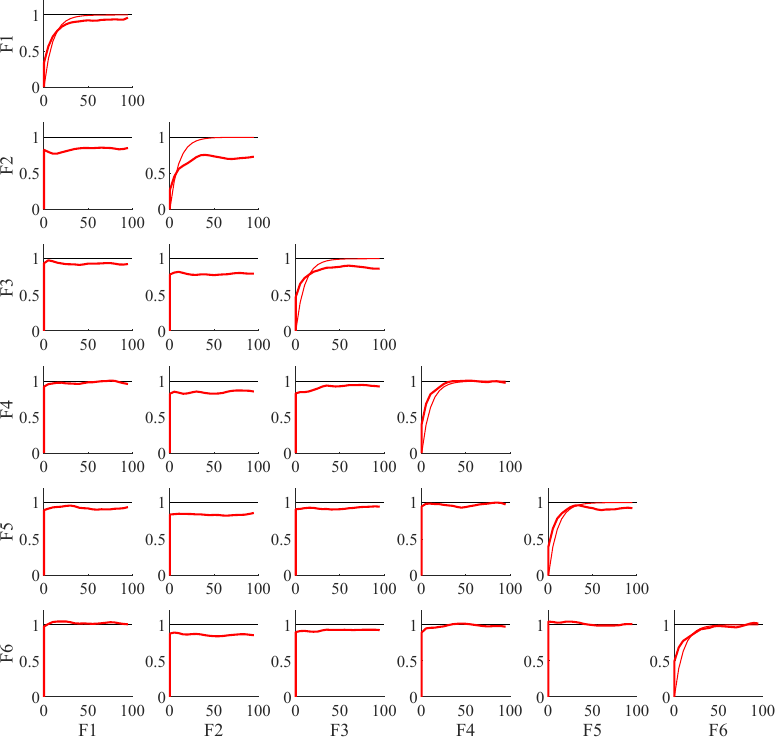}
	\caption{Experimental variogram of the gaussian factors, and the final model used.}
	\label{variogram}
\end{figure*}

Variogram analysis and calibration is the weakest point of the methodology. A first complication is that de-correlation breaks the marginal guassianity on the factors $ \textbf{Y}=\textbf{L}^{-1}\textbf{Z} $, suggesting that the assumption of multi-gaussianity on $ \textbf{Z} $ is not a perfect model at every location. As a consequence, the experimental variance on the factors $ \textbf{Y} $ do not attains the value of 1, although it gets close for a couple of factors. This fact can be seen on the sill of the experimental  variograms in Fig. \ref{variogram}. However, one-structured exponential variogram with 10 m of range is fixed as a final model, fitting relatively well for most of direct variograms. Cross variograms show low correlation among variables, as expected, however the sill do not attains 1 in some of the cases, in the same fashion as previously described. This last effect was not considered nor included. 

\subsection{Results}

\begin{figure}[h!]
	\centering
	\includegraphics[width=0.48\textwidth,trim={0 0cm 0.0cm 0cm},clip]{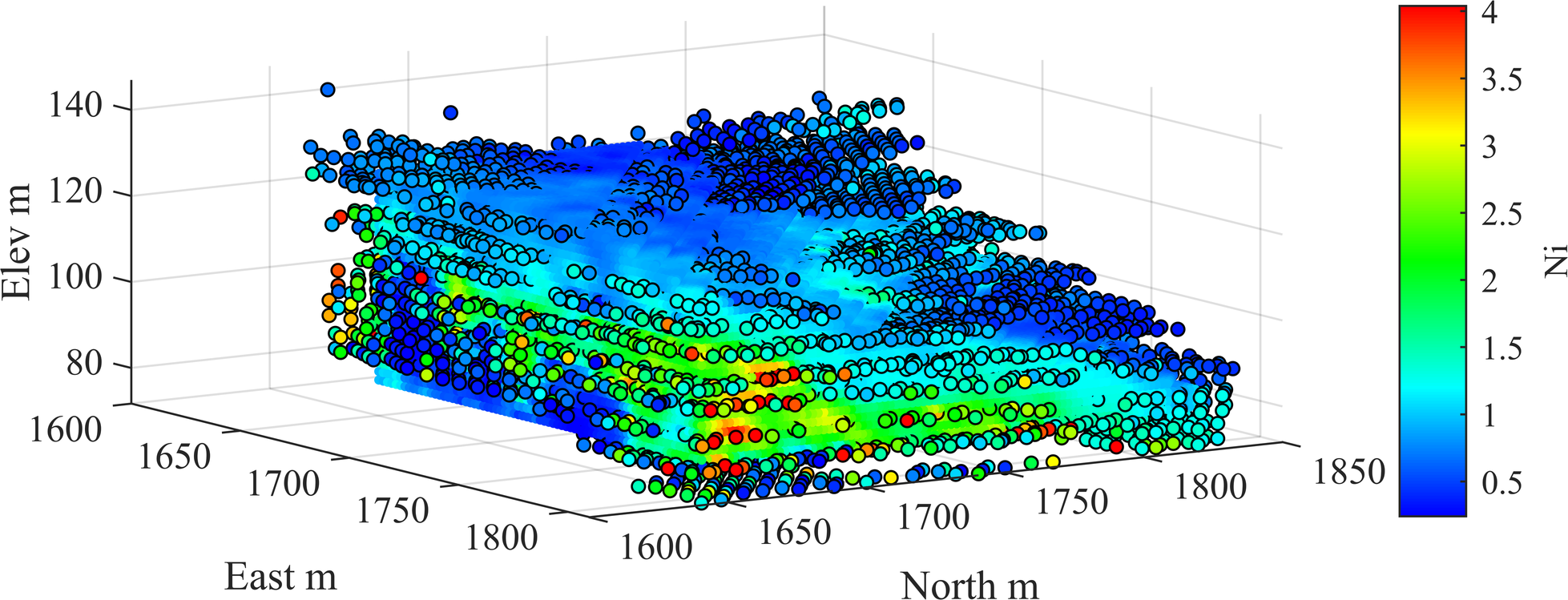}
	\includegraphics[width=0.48\textwidth,trim={0 0cm 0.0cm 0cm},clip]{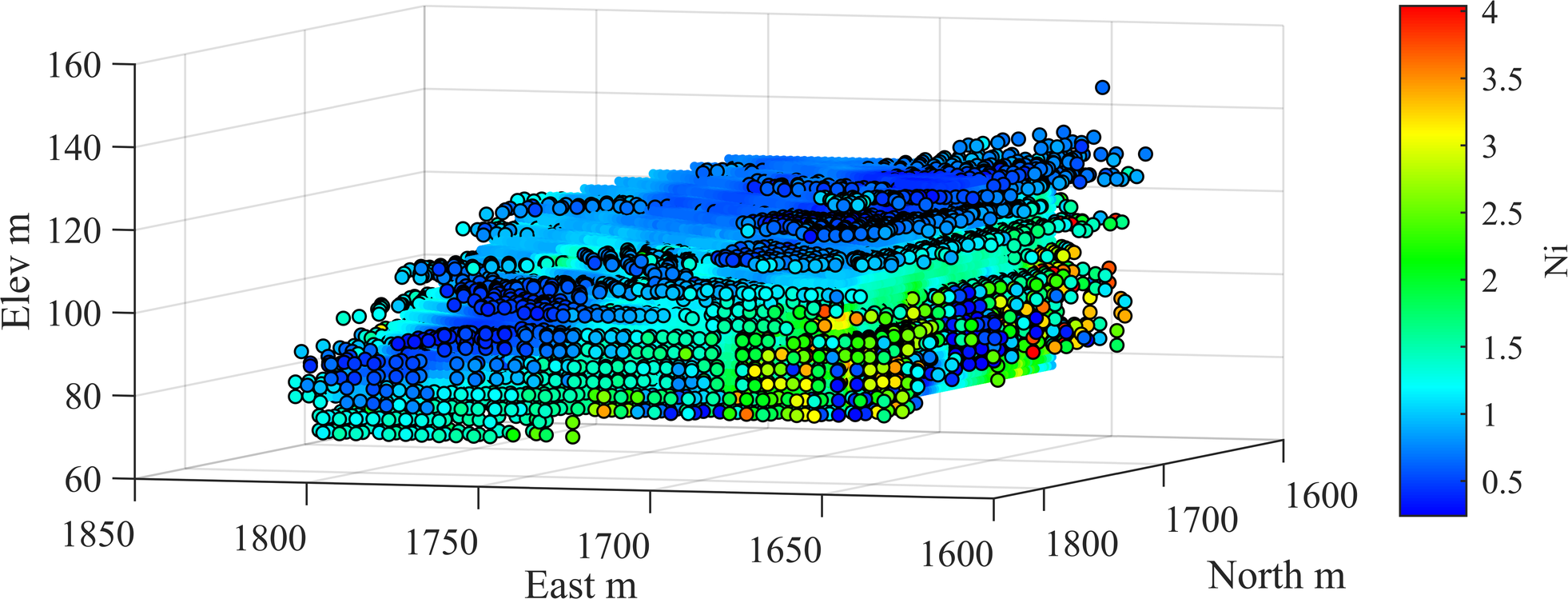}
	\caption{Grid used for the study showing the estimated mean and sampling data, for the Nickel case.}
	\label{grid}
\end{figure}

\begin{figure}[h!]
	\centering
	\includegraphics[width=0.48\textwidth,trim={0 0cm 0.0cm 0cm},clip]{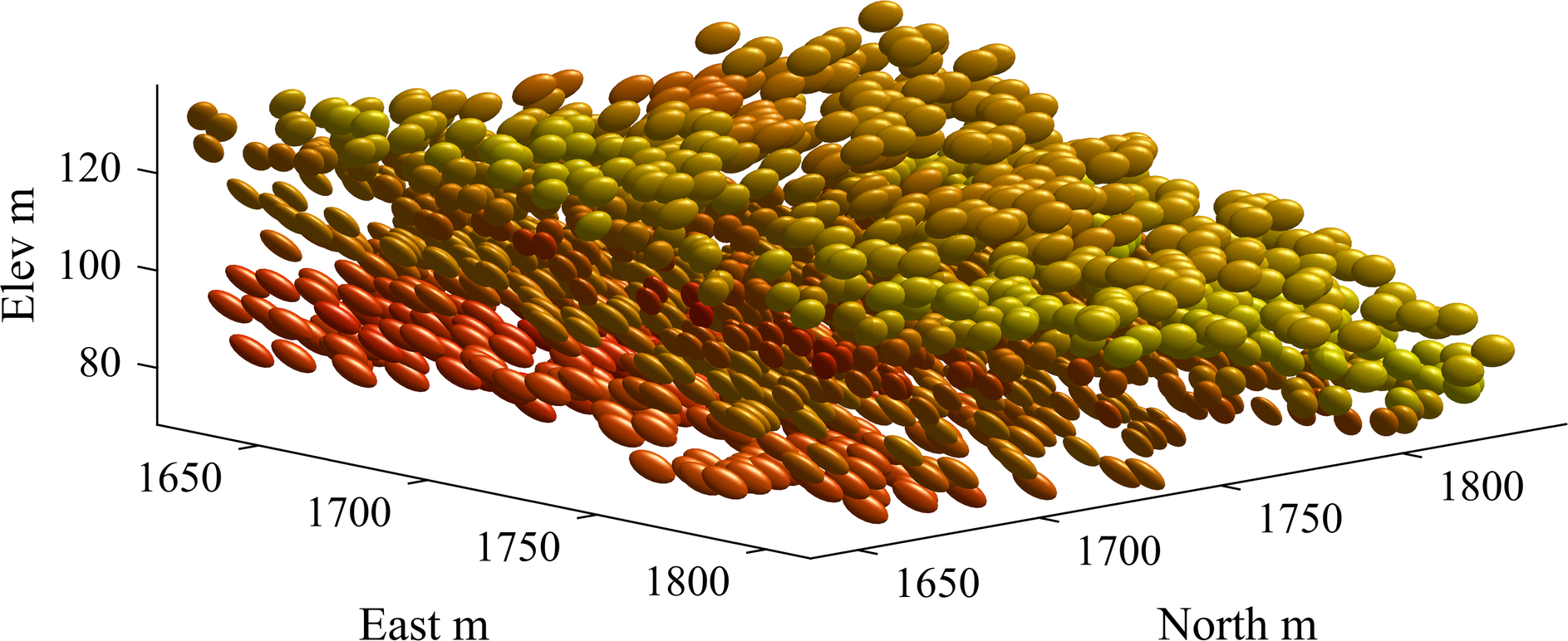}
	\includegraphics[width=0.48\textwidth,trim={0 0cm 0.0cm 0cm},clip]{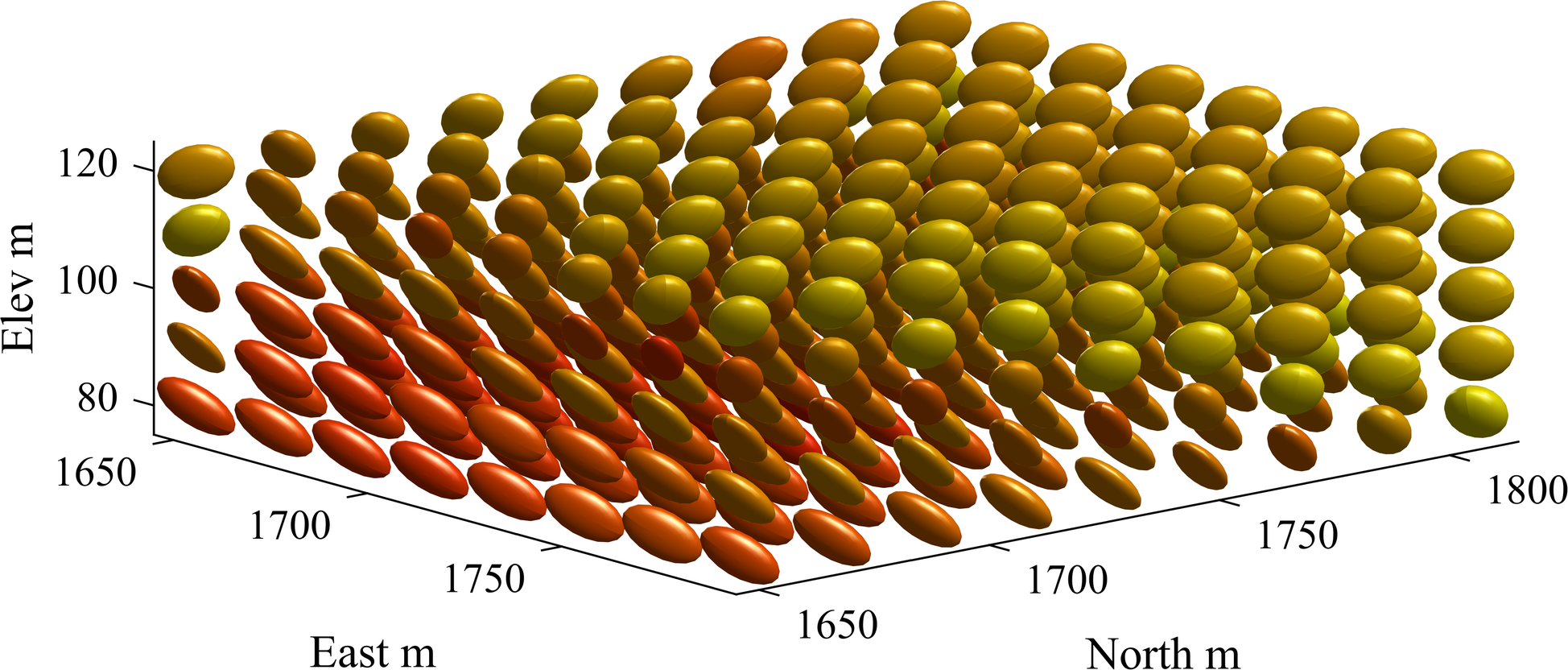}
	\includegraphics[width=0.15\textwidth,trim={0 0cm 0.0cm 0cm},clip]{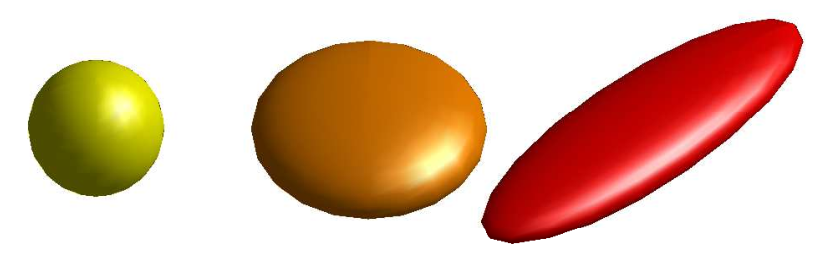}
	\caption{Estimated correlation matrices at sample positions (correlation of Ni, Fe and Mg), represented by ellipses (left). Interpolation of correlation matrices on a regular grid (right).The color of ellipsoids is related to their anisotropy (bottom). From left to right: isotropic tensor, planar tensor (flat ellipsoid) ($ \lambda_1 \simeq \lambda_2 > \lambda_3 $), elongated ellipsoid ($ \lambda_1 \gg \lambda_2 \geq \lambda_3 $)}
	\label{gridelli}
\end{figure}

\begin{figure*}[h!]
	\centering
	\includegraphics[width=0.39\textwidth,trim={0 0cm 0.0cm 0cm},clip]{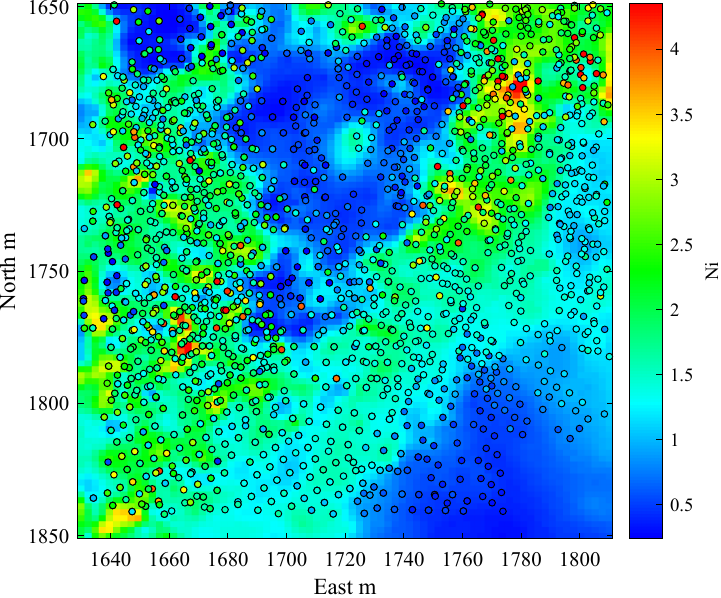}
	\includegraphics[width=0.39\textwidth,trim={0 0cm 0.0cm 0cm},clip]{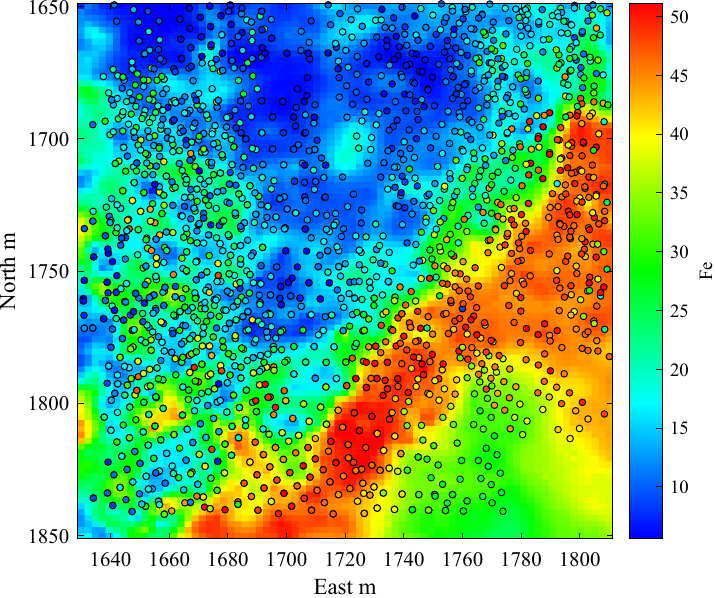}
	\includegraphics[width=0.39\textwidth,trim={0 0cm 0.0cm 0cm},clip]{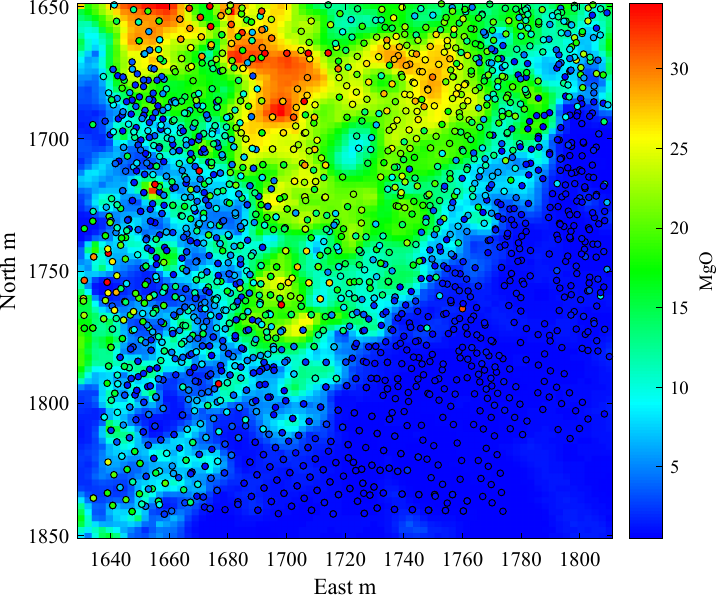}
	\includegraphics[width=0.39\textwidth,trim={0 0cm 0.0cm 0cm},clip]{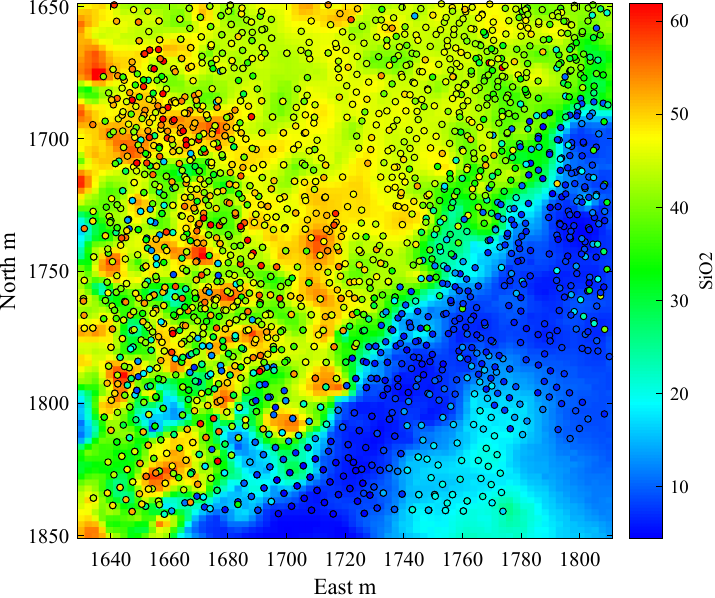}
	\includegraphics[width=0.39\textwidth,trim={0 0cm 0.0cm 0cm},clip]{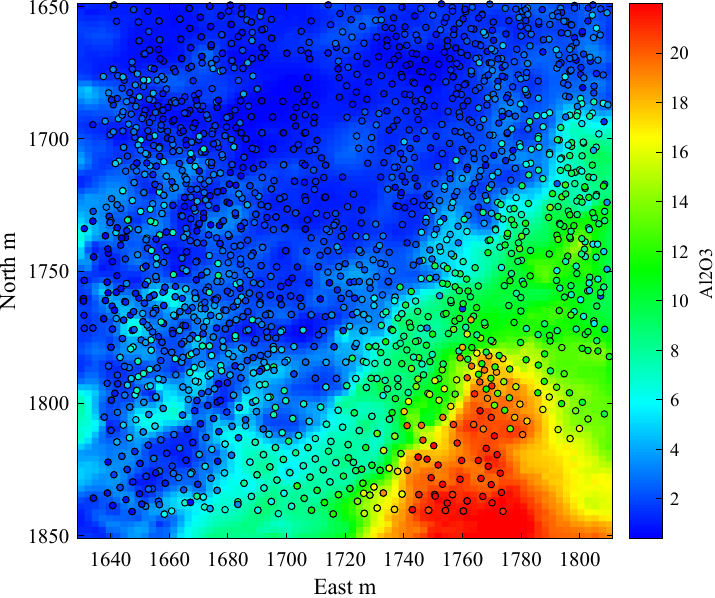}
	\includegraphics[width=0.39\textwidth,trim={0 0cm 0.0cm 0cm},clip]{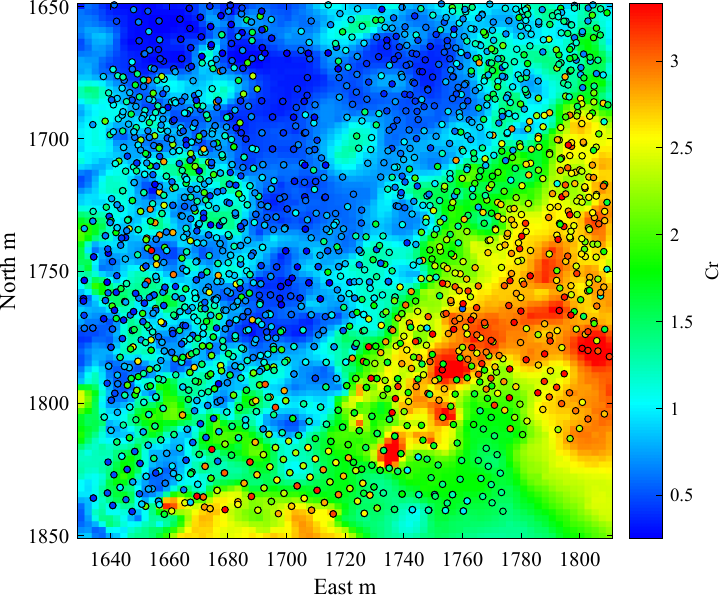}
	\caption{Plan view showing the estimated mean of the simulations at level 95 m and sampling data.}
	\label{mean}
\end{figure*}

Once the only variogram formulae is derived, one can establish the simply to work individually on each of the factor. A initial grid with mesh dimension of 2×2×2 (in meters) with mesh size of 75, 90 and
25 along east, north, and elevation coordinates is considered.
We proceed to generate 1000 geostatistical simulations by using turning band algorithm \citep{Chiles} (picking 1200 directions). The neighborhood is selected as moving and the parameters for the range of search neighborhood are set to 100 m with up
to 25 number of data and without considering octants. This number is chosen arbitrarily as the scope of this study is mainly focused on the examination of uncertainty, being the  number of data chosen
for simulation not relevant for the study. The simulated factors are later correlated according to the estimated correlation, interpolated by ordinary kriging (in order to get weights adding 1) at each location of the grid, by using the same variogram model as for the factors, and then back-transformed from gaussian values and from log-ratios into the raw distribution. The filtered grid for the mean of the simulations in the Nickel case, excluding the nodes far from sample data, is shown in Fig. \ref{grid}. The estimated correlation at sample locations and the interpolation on the regular grid is shown in Fig. \ref{gridelli}. The produced maps showing the mean of the simulations, at level 95 m, is given in Fig. \ref{mean}, for the six back-transformed cross-correlated variables.
The results reproduce cross-correlation trends in the maps. For instance, there is a strong negative correlation
between Fe and MgO, which can be corroborated from visual inspection.

%\begin{figure}[h!]
%	\centering
%	\includegraphics[width=0.49\textwidth,trim={0 0cm 0.0cm 0cm},clip]{./fotos/loc_sim1.pdf-1.png}
%	\includegraphics[width=0.49\textwidth,trim={0 0cm 0.0cm 0cm},clip]{./fotos/loc_sim2.pdf-1.png}
%	\includegraphics[width=0.49\textwidth,trim={0 0cm 0.0cm 0cm},clip]{./fotos/loc_sim3.pdf-1.png}
%	\includegraphics[width=0.49\textwidth,trim={0 0cm 0.0cm 0cm},clip]{./fotos/loc_sim4.pdf-1.png}
%	\includegraphics[width=0.49\textwidth,trim={0 0cm 0.0cm 0cm},clip]{./fotos/loc_sim5.pdf-1.png}
%	\includegraphics[width=0.49\textwidth,trim={0 0cm 0.0cm 0cm},clip]{./fotos/loc_sim6.pdf-1.png}
%	\caption{Plan view showing one of the simulations at level 95 m and sampling data.}
%	\label{sim}
%\end{figure}

\begin{figure}[h!]
	\centering
	\includegraphics[width=0.32\textwidth,trim={0 0cm 0.0cm 0cm},clip]{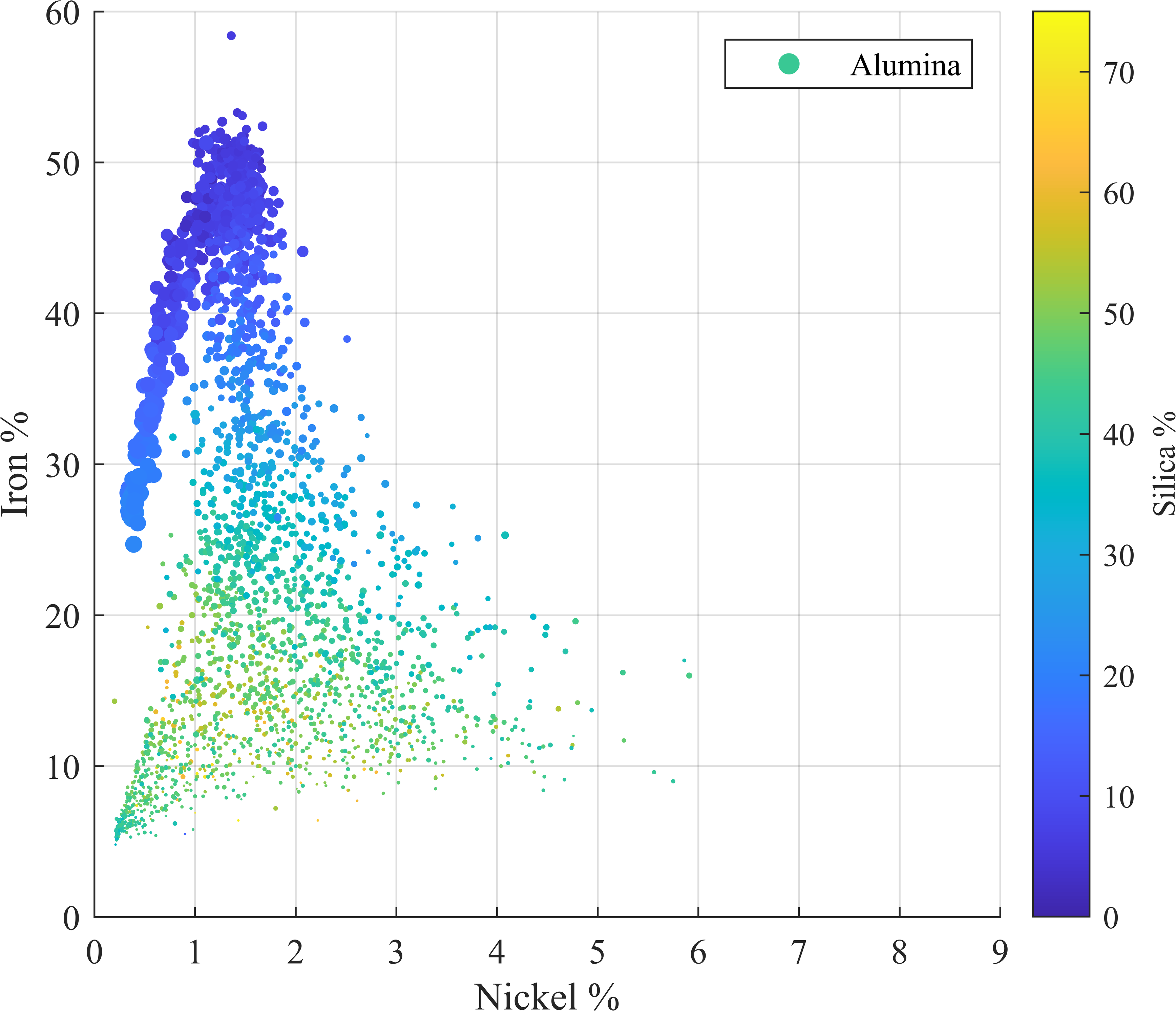}
	\includegraphics[width=0.32\textwidth,trim={0 0cm 0.0cm 0cm},clip]{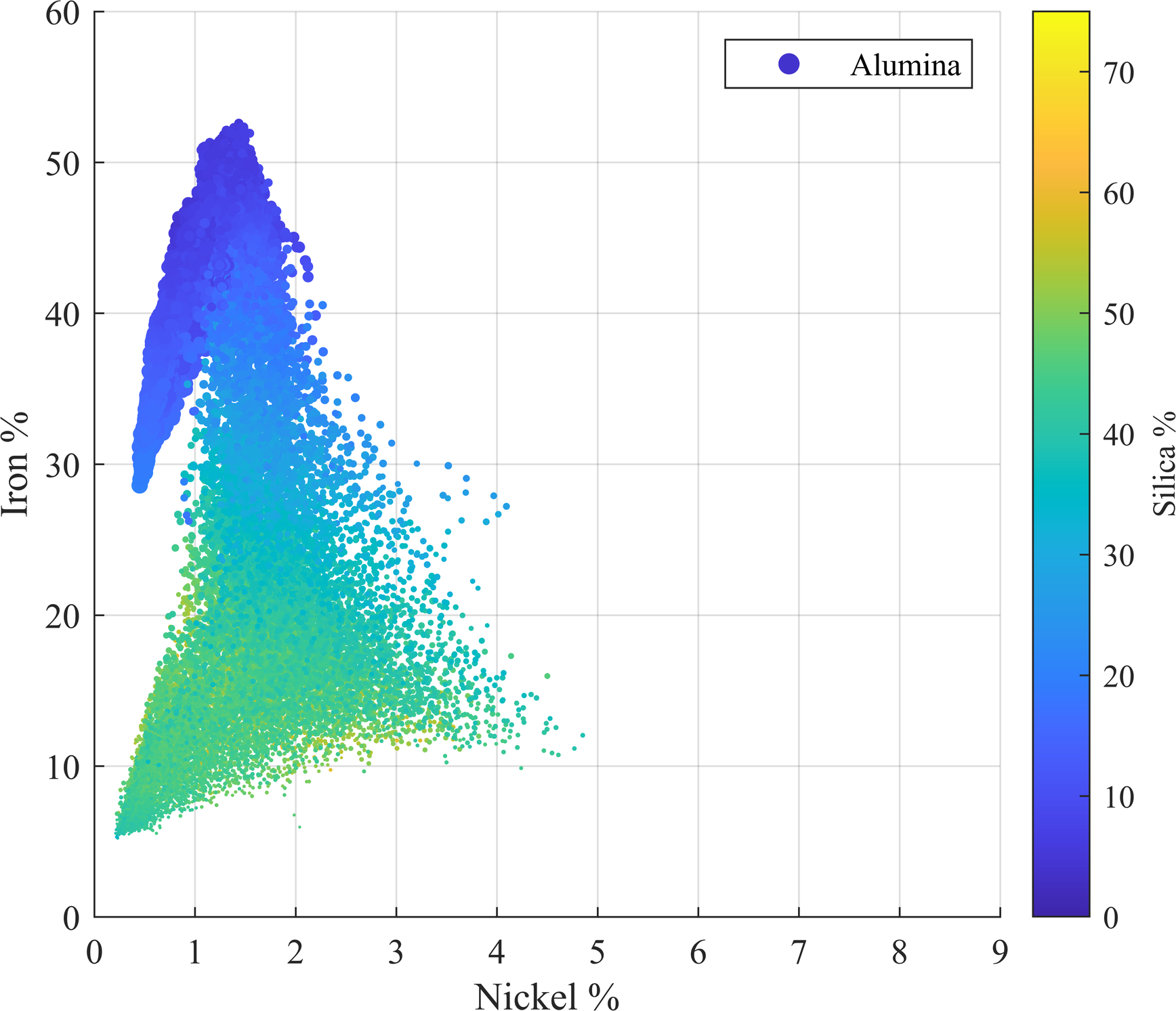}
	\includegraphics[width=0.32\textwidth,trim={0 0cm 0.0cm 0cm},clip]{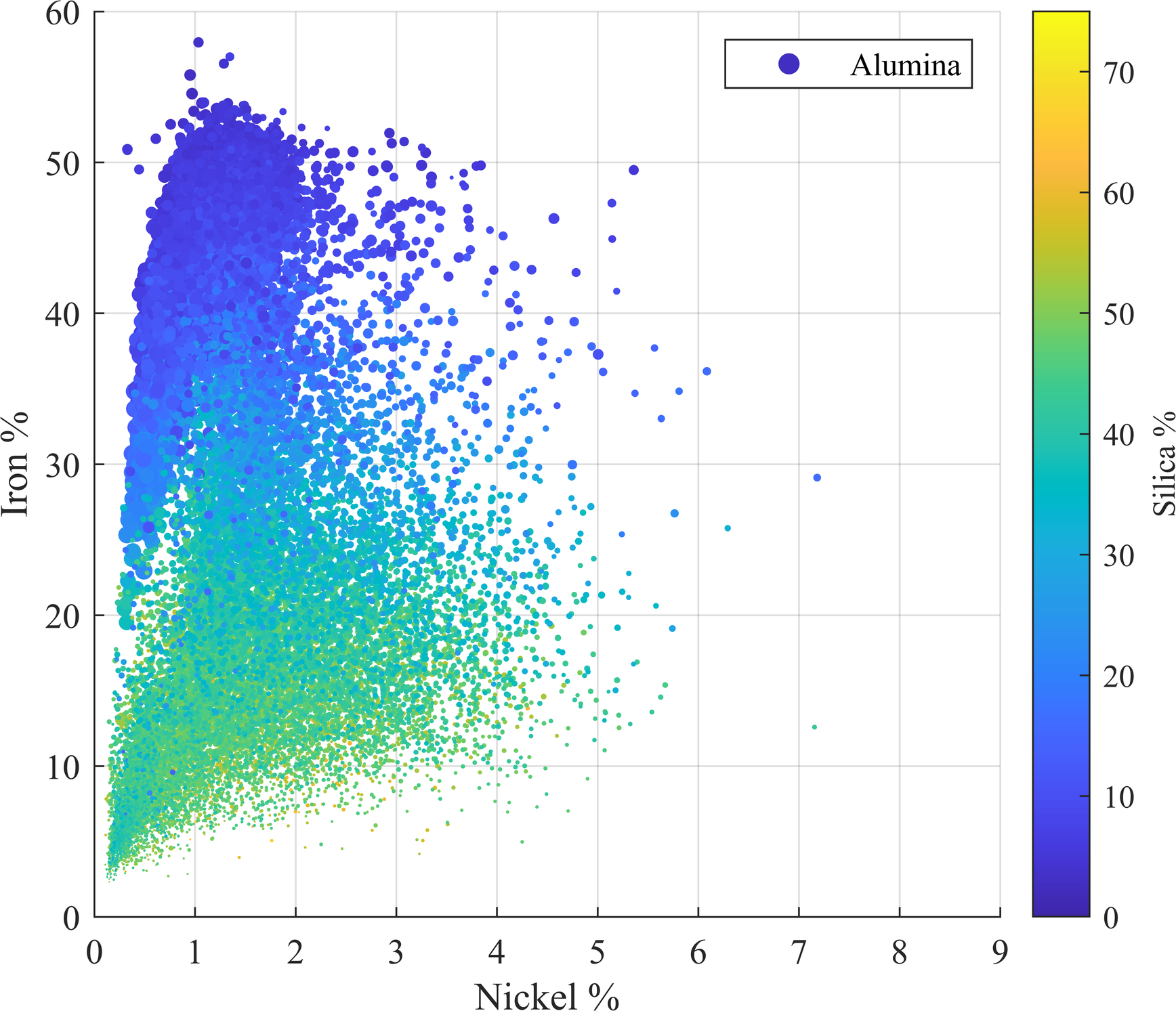}
	\caption{Scatter plot for sampling data, the mean on the grid values and one simulation respectively from the left, around level 95 m.}
	\label{scatt2}
\end{figure}

\begin{figure*}[h!]
	\centering
	\includegraphics[width=0.8\textwidth,trim={0 0cm 0.0cm 0cm},clip]{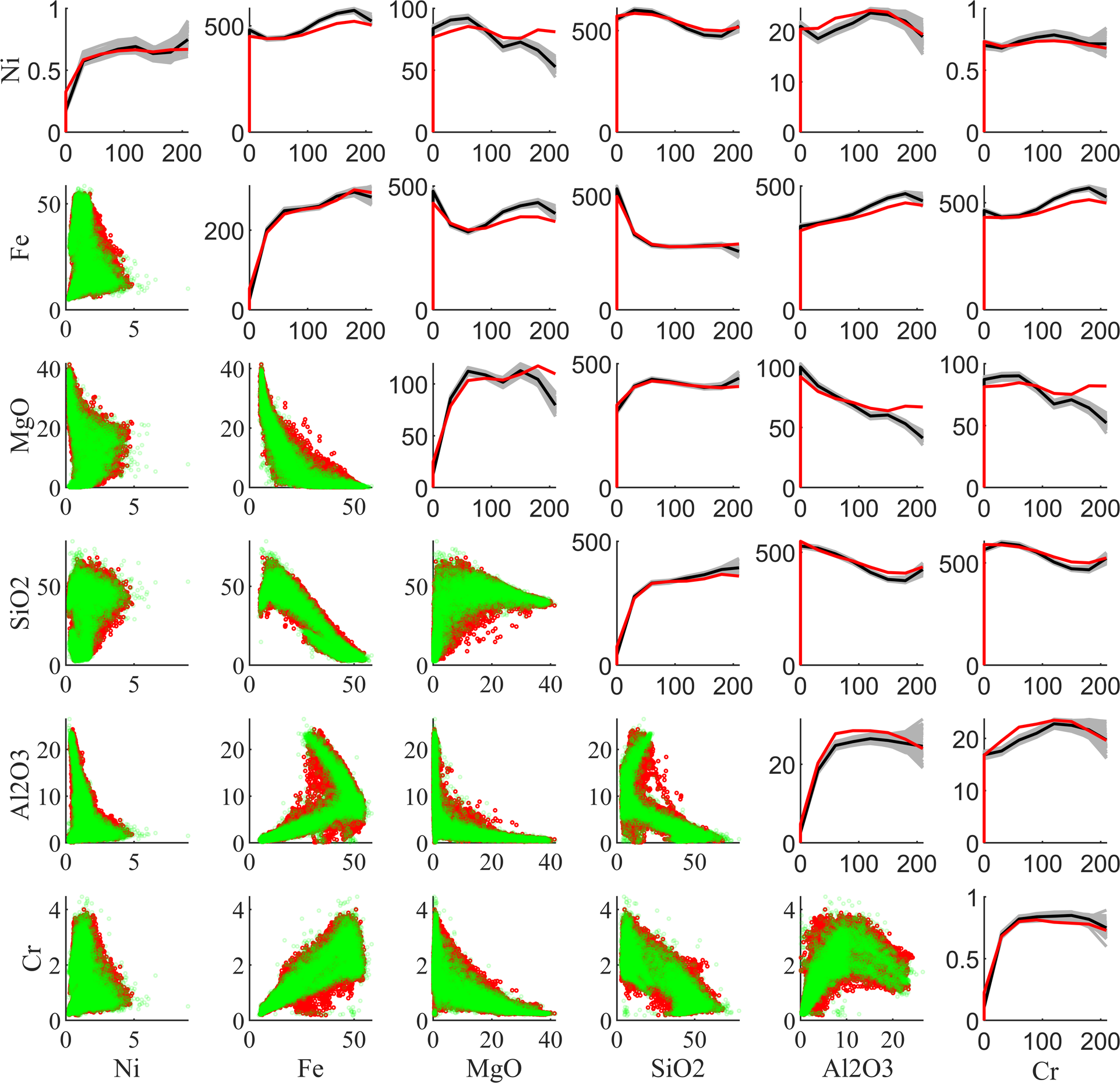}
	\caption{Scatter plots and variograms comparing ground truth and mean of simulations. Variograms from simulations are shown in light gray lines, the mean of the variogram on black line, and the variogram of the ground truth on red line. The scatter plot showing bi-variate relations is shown on green dots, superposed to the mean of simulations on red dots.}
	\label{bivar}
\end{figure*}

The non-linear behavior among variables is well reproduced. This is shown on Fig. \ref{scatt2} in the case of the mean of the simulations and for one of them, around level 95 m. Scatter plots showing all bi-variate relations for the mean of the simulations are shown in Fig. \ref{bivar}, together with the results of the variography. Variograms are well reproduced, besides the issues commented previously. It is quite impressive how well-fitted are most of the direct and cross variograms, given the fact that only one variogram was considered for the purpose of the presented methodology.

In order to test the predictability and the uncertainty assessment capabilities of the methodology, we bring back the testing data leaved out from the first part of the case study. Each testing data was linked to the closest node on the grid, for retaining only the data within less than 2.5 m in distance to the corresponding node, to avoid distortions on results, leaving finally 366 samples to be considered from the initial 500 in an uncertainty analysis.

\begin{figure*}[h!]
	\centering
	\includegraphics[width=0.8\textwidth,trim={0 0cm 0.0cm 0cm},clip]{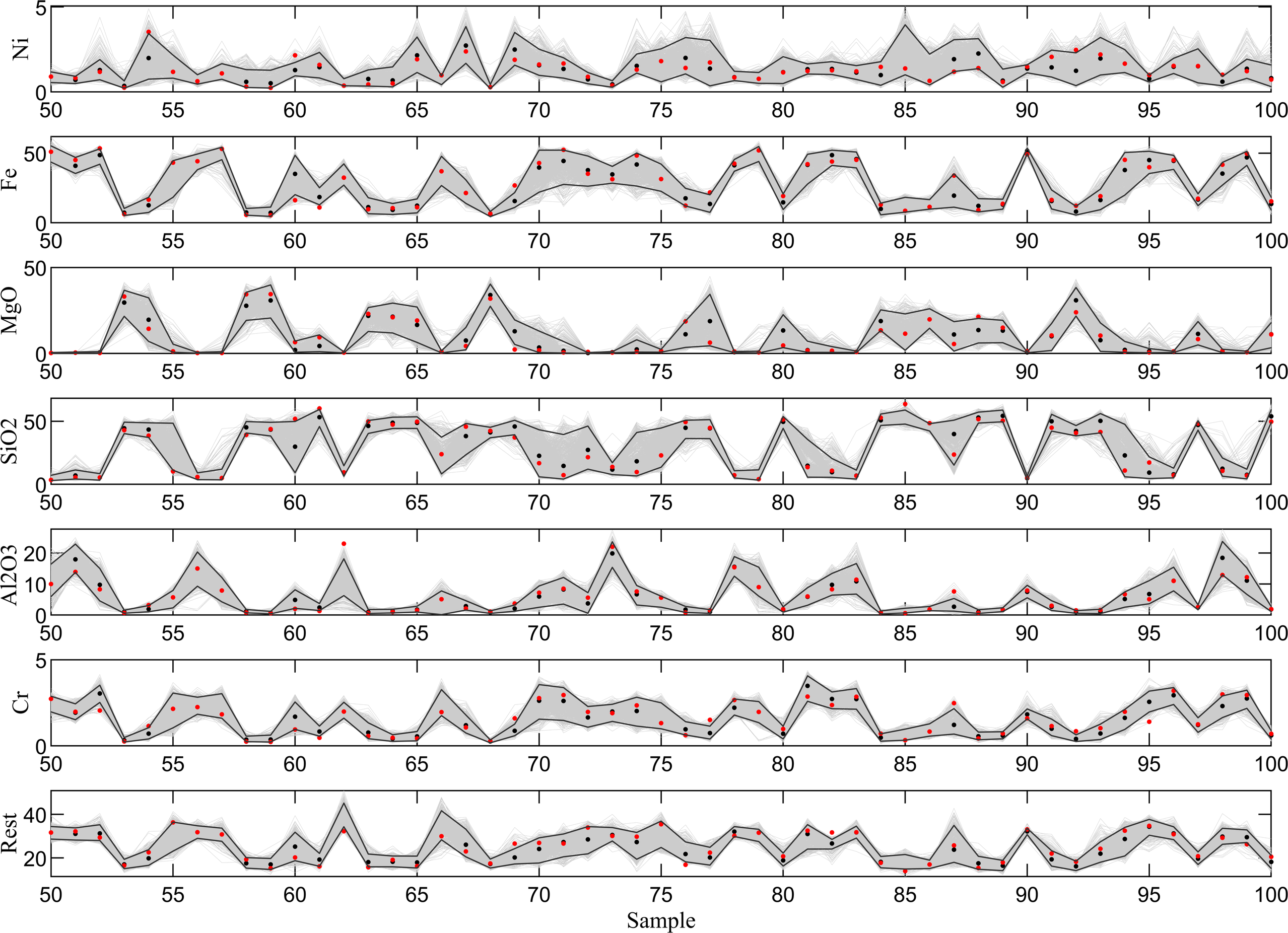}
	\caption{Uncertainty assessment for 50 samples taken from testing data, showing 1000 simulations in gray lines, the lower 5\% and the upper 95\% percentile as confidence boundary in  black lines, the estimated mean in black dots and, in red dots, the ground truth.}
	\label{samples2}
\end{figure*}

The resulting pdfs from the simulations are obtained and shown in Fig. \ref{samples2} for 50 samples. We pick this small window to inspect results in detail. The realizations are display of light gray lines, and the mean estimation of the simulation (in black dots) is shown for the different seven variables, at the 366 samples. Red dots represent the true grade of the samples. A 5\% and 95\% percentile lines are displayed in black lines to give a 90\% confidence area.

%\begin{figure}[h!]
%	\centering
%	\includegraphics[width=0.99\textwidth,trim={0 0cm 0.0cm 0cm},clip]{./fotos/samples_pred.png}
%	\caption{Uncertainty assessment for testing data, showing 1000 simulations in gray lines, the lower 5\% and the upper 95\% percentile as confidence boundary in  black lines, the estimated mean in black dots and, in red dots, the ground truth.}
%	\label{samples1}
%\end{figure}

\begin{figure*}[h!]
	\centering
	\includegraphics[width=0.39\textwidth,trim={0 0cm 0.0cm 0cm},clip]{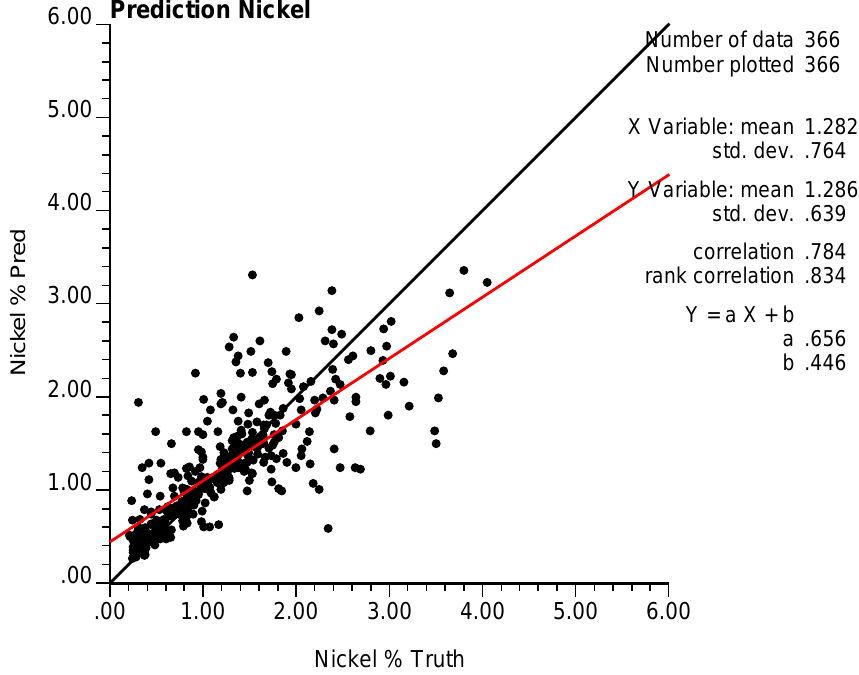}
	\includegraphics[width=0.39\textwidth,trim={0 0cm 0.0cm 0cm},clip]{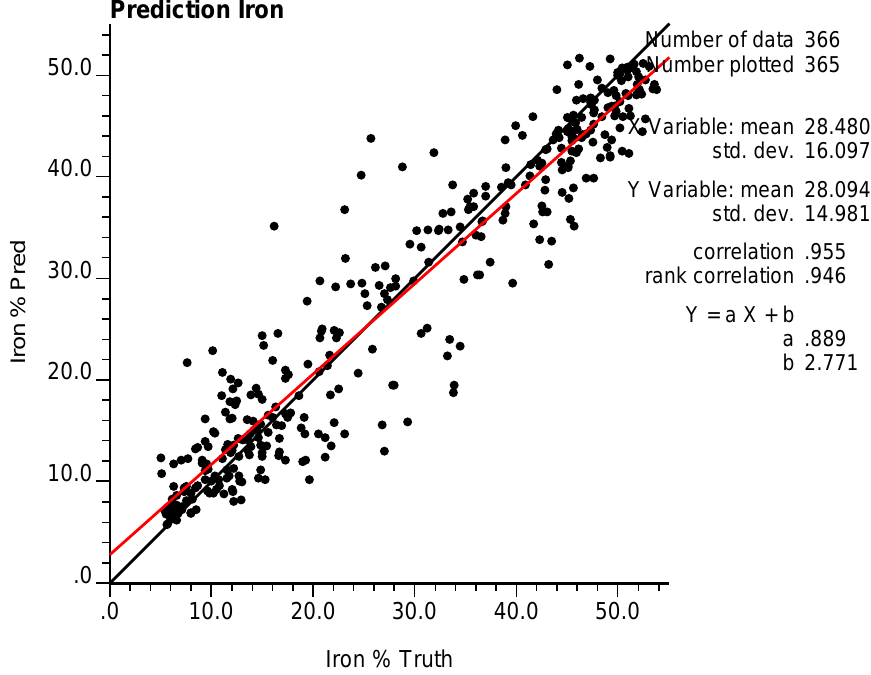}
	\includegraphics[width=0.39\textwidth,trim={0 0cm 0.0cm 0cm},clip]{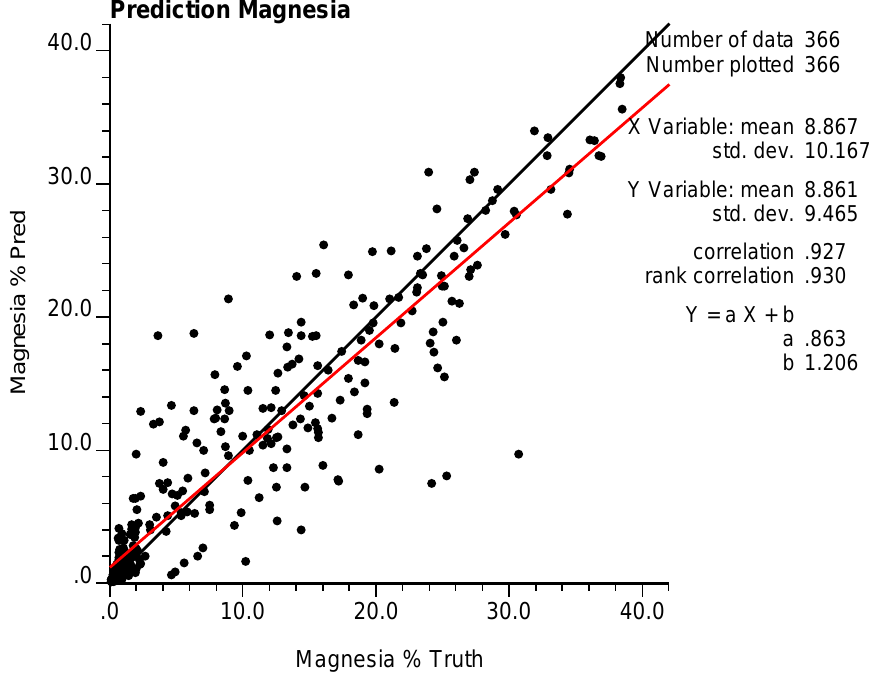}
	\includegraphics[width=0.39\textwidth,trim={0 0cm 0.0cm 0cm},clip]{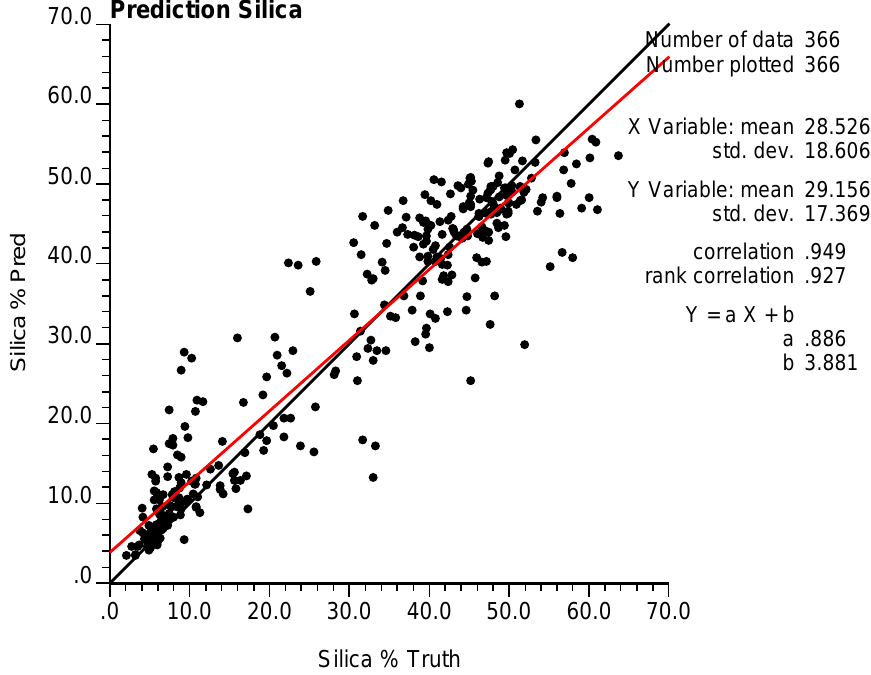}
	\includegraphics[width=0.39\textwidth,trim={0 0cm 0.0cm 0cm},clip]{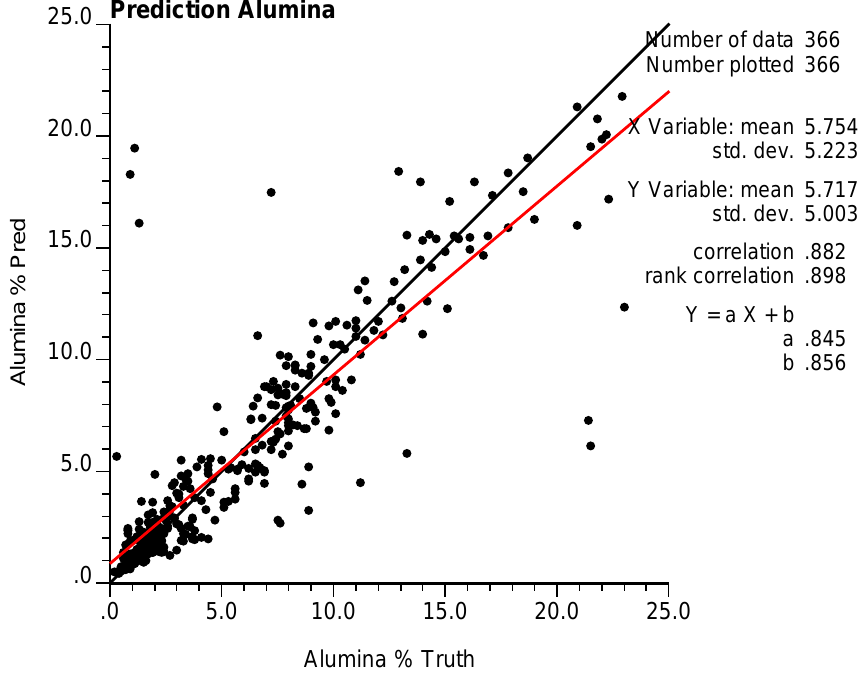}
	\includegraphics[width=0.39\textwidth,trim={0 0cm 0.0cm 0cm},clip]{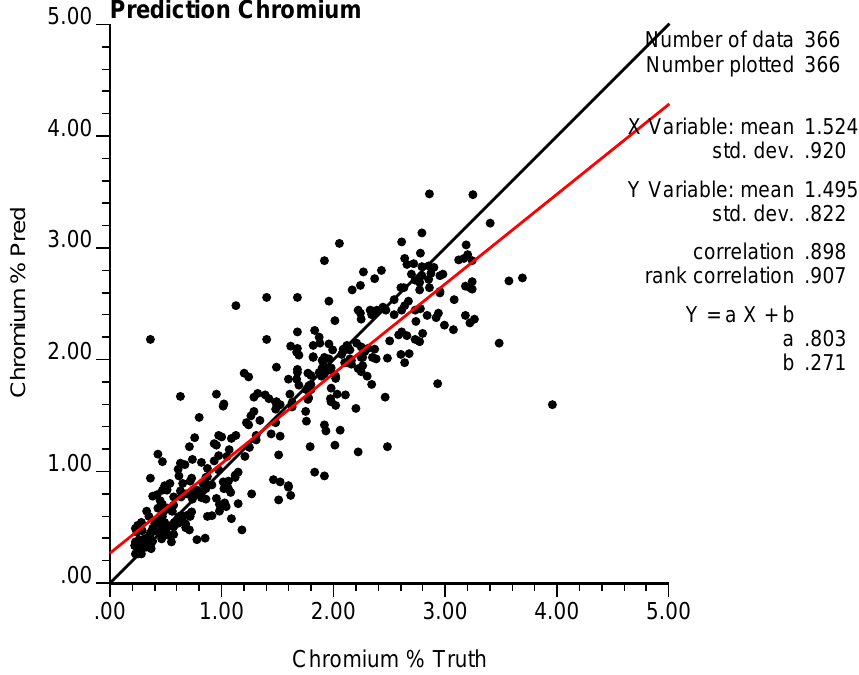}
	\caption{Scatter plots comparing the estimated mean of the simulations on locations closed to testing data with the ground truth.}
	\label{scatt1}
\end{figure*}

Figure \ref{scatt1} shows the scatter plots comparing the estimated mean of volumes versus their ground truth value, as the mean a value is often taken as a predictor for the real value. Low bias on the prediction and high correlation values are obtained, varying from a lowest value of 0.78 (in the case of Nickel) to 0.95 (in the case of Iron).

\begin{figure}[h!]
	\centering
	\includegraphics[width=0.49\textwidth,trim={0 0cm 0.0cm 0cm},clip]{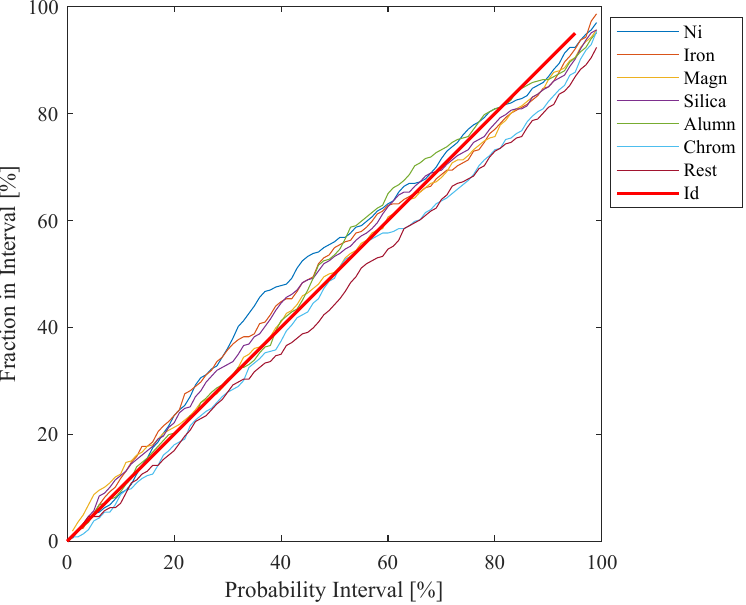}
	\caption{Accuracy plot, which calculates the proportion of locations where the true value falls within symmetric $ p $-probability interval.}
	\label{acc}
\end{figure}

The main advantage of simulating is that we can validate if the decision made on previous steps was correct. The validation is completed with the generation of an accuracy plot to check that the uncertainty given by the pdfs effectively represents the experimental frequencies on the ground truth of testing data (Fig. \ref{acc}).

\begin{figure}[h!]
	\centering
	\includegraphics[width=0.32\textwidth,trim={0 0cm 0.0cm 0cm},clip]{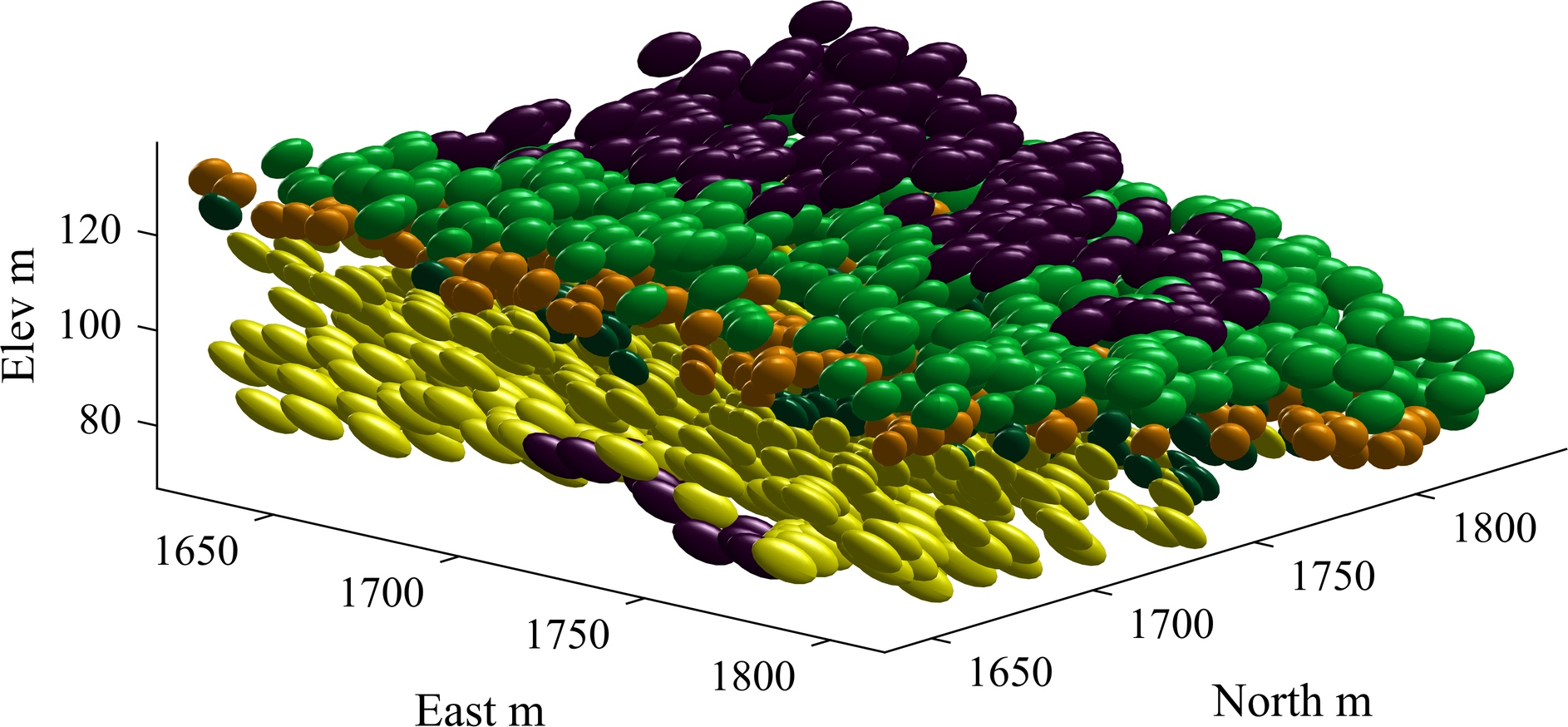}
	\includegraphics[width=0.32\textwidth,trim={0 0cm 0.0cm 0cm},clip]{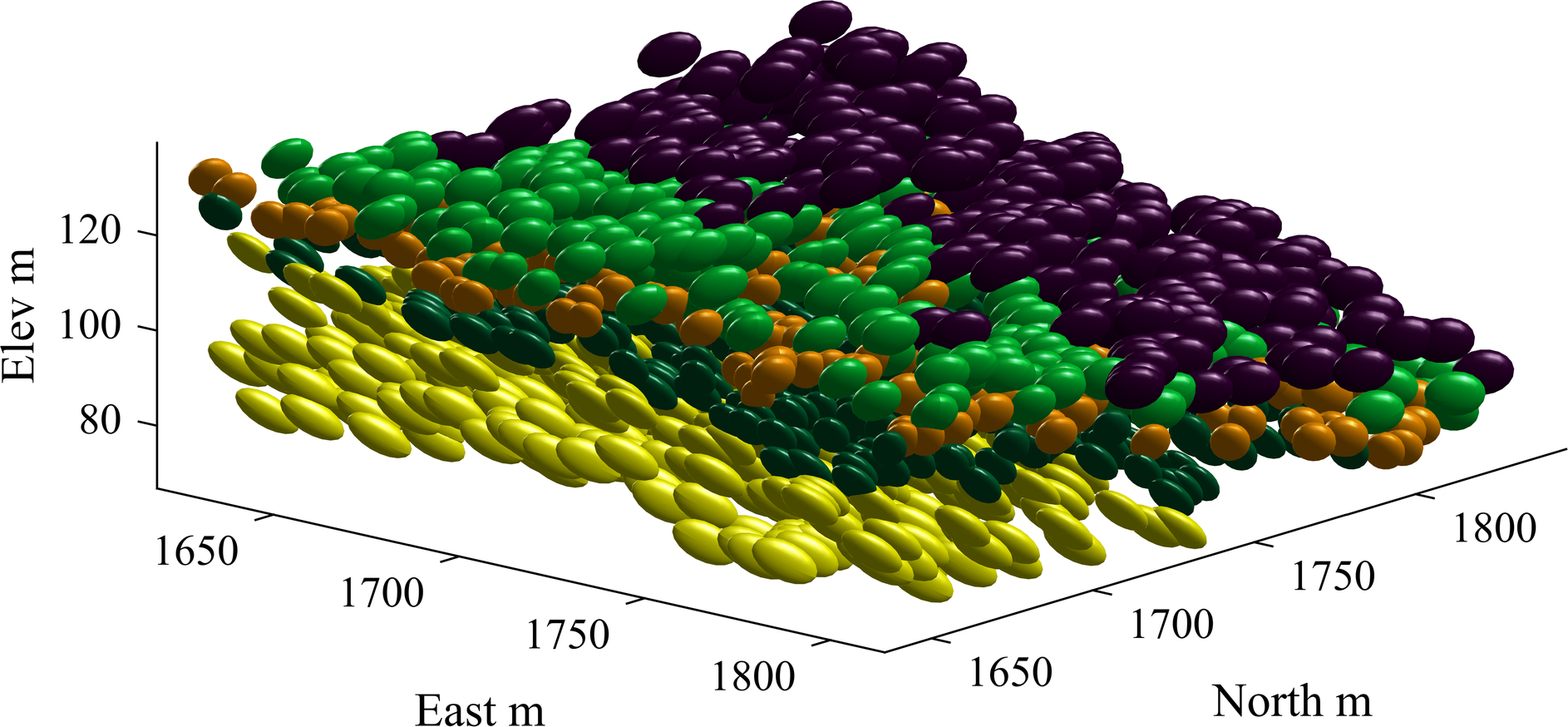}
	\includegraphics[width=0.32\textwidth,trim={0 0cm 0.0cm 0cm},clip]{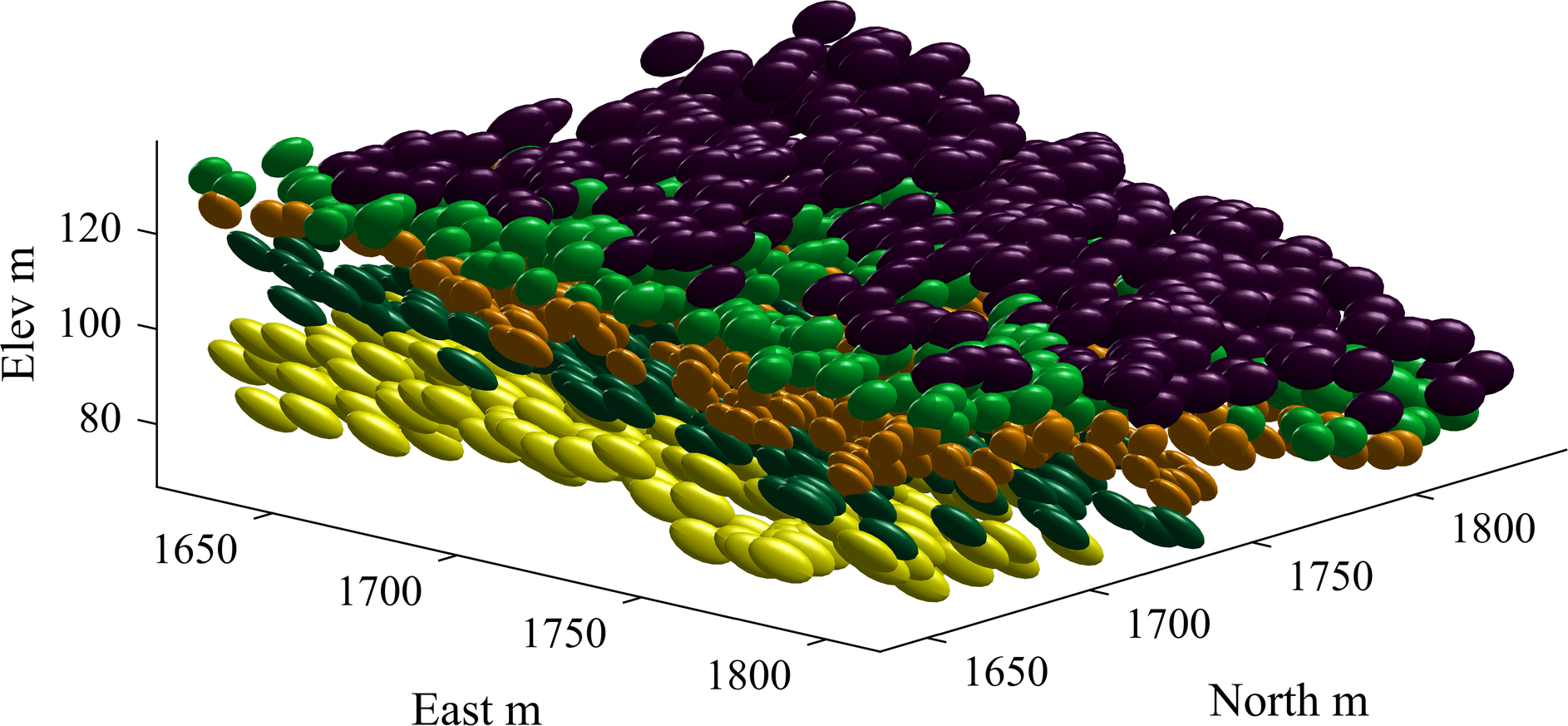}
	\caption{Isometric view of ellipses at samplig location showing the progress on clustering when using $ K $-means algorithm on sampling data, for $ K=5 $, and for iterations 1, 3, and 30 respectively from the top.}
	\label{clust2}
\end{figure}

\begin{figure}[h!]
	\centering
	\includegraphics[width=0.32\textwidth,trim={0 0cm 0.0cm 0cm},clip]{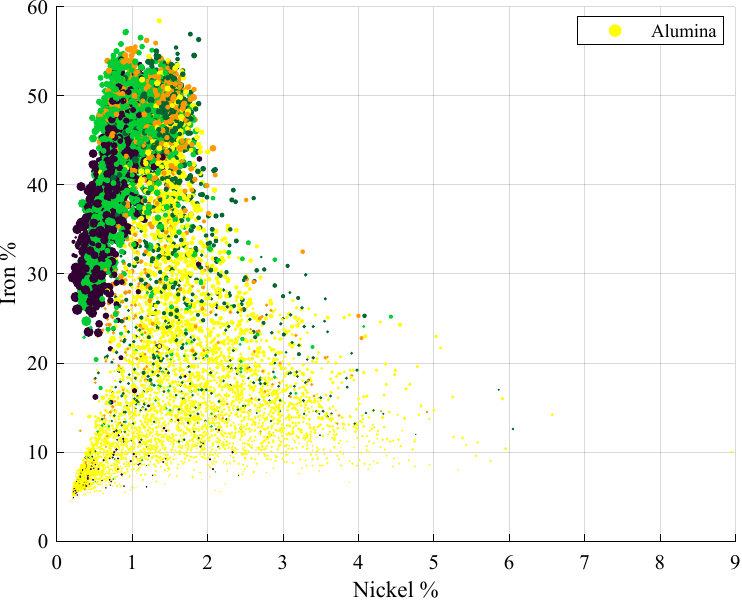}
	\includegraphics[width=0.32\textwidth,trim={0 0cm 0.0cm 0cm},clip]{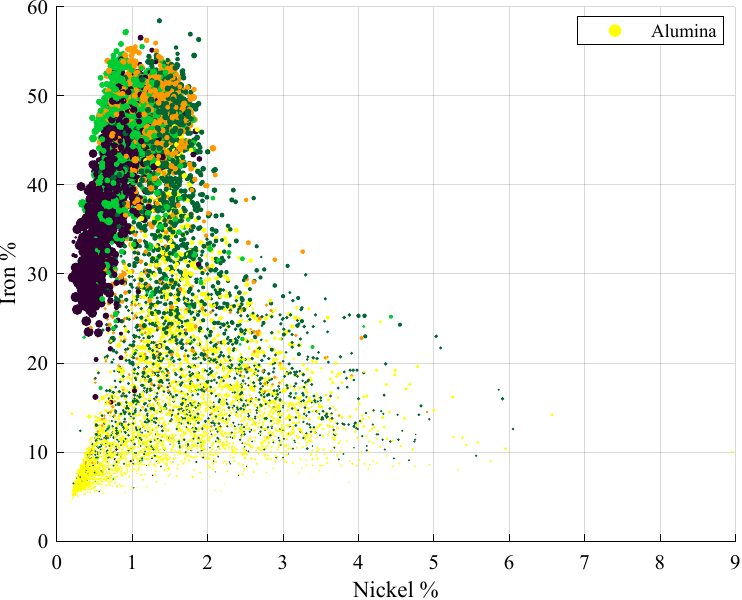}
	\includegraphics[width=0.32\textwidth,trim={0 0cm 0.0cm 0cm},clip]{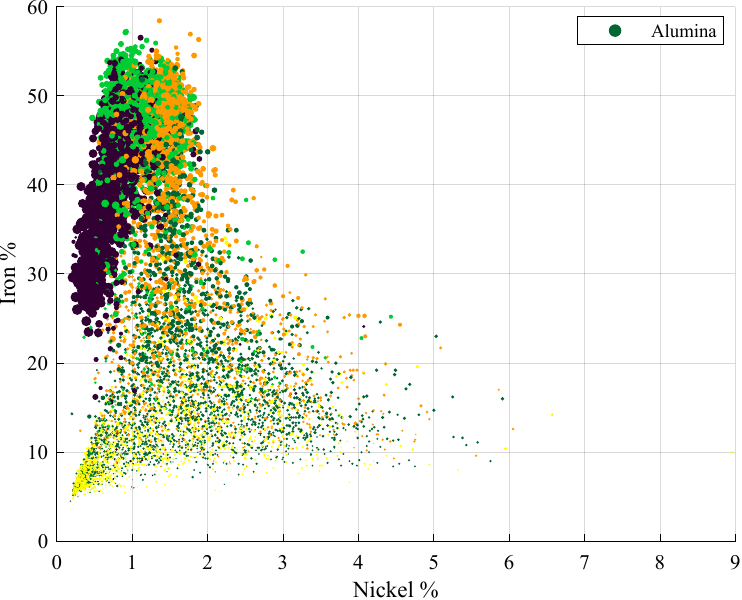}
	\caption{Scatter showing the progress on clustering when using $ K $-means algorithm on sampling data, for $ K=5 $, and for iterations 1, 3, and 30 respectively from the top.}
	\label{clust}
\end{figure}

Finally, in Figs. \ref{clust2} and \ref{clust} , we present the clustering algorithm results, with $ K=5 $, showing high consistency between the clusters obtained at the last stage and the geological units tagged in the database, respectively. The algorithm even seems to capture the directions of continuity shown on the grades by the data. Spatial continuity in the units provided by the algorithm and, besides starting with some artifacts at the initial step, rapidly on iteration number three one can anticipate the definitive zones, demonstrating to be a promising tool at the moment of delimitation of stationary units. 

\section{Conclusions}
\label{Conclusions}

%Linear interpolation of geo-related variables is important to different aspects of mining engineering such as long- and short-term mine planning. In the case of complex ore deposits, the issue of estimation of variables in the region is questionable and somehow tedious when using state-of-the-art techniques. Due to this fact, modeling those complications in the co-spatial behavior of the grades in multi-element deposits prompts one to employ enhanced geostatistical techniques, since simple traditional approaches for this purpose are incapable of reproduce those complexities such as non-linearity, heteroscedasticity, and geological constraint that frequently exist among the variables of interest.
We have shown how multi-variate data can be modeled and understood as a RF lying on a correlation manifold, on where one can map every data sample into this topological space. By using this tool, two applications follow: first, the interpolation of the different known correlation matrices throughout the domain with the purpose of reproducing the non-linear multivariate features of data; and second, an application which deals with the problem of clustering of multivariate data.

As a summary, a conceptually simple and novel methodology has been proposed to account for non-linearity in multivariate data, with reasonably good results that reproduce the multivariate behavior. By integrating some basic aspects of Riemannian geometry and the well known machinery for handling SPD matrices into the geostatistical setting, we gain enough flexibility to reproduce the mentioned complex multivariate behavior and, at the same time, serves to improve our understanding in the geological data. Implementing interpolation of correlation matrices for carrying the local linear multivariate relationships is a key step in good reproduction of data behavior.

Among the limitations, we mention that the proposed methodology
only works with enough data to estimate the correlation locally. As with other methodologies that try to handle non-stationarity, when limited data is available, it is better to simplify the problem and assume stationarity on the data, as calibration of hyperparameters, such as the correlation matrix at the different locations, may become hard to obtain. The variography becomes theoretically challenging to handle and interpret under the assumption of different underlying structures as well, since working with different models of spatial continuity for the different structures is a tricky decision to make as a ``rotation'' of the structures is a free parameter and a valid model that also fits the spatial correlation among variables. A third issue is that the definition of stationary geological domains beforehand replaces the presented methodology. If the multivariate behavior changes ``continuously'', the proposed methodology may be a promising approach for handling non-stationary.

As part of future research tasks, we propose developing a synthetic study to fully understand some crucial details of the methodology, such as the sensibility analysis of the main parameters involved and the impact to adjustments. Among these parameters, that may be critical for the method, we mention the local neighborhood from which the correlation matrix is obtained, at sample locations. How sensitive is the estimation for this correlation matrix to the number of data used, and the possibility of using a variable size in the amount of data are some of the open questions to be answered. Further efforts has to be made to give meaning to the variogram modeling step when including different models of spatial structure and the effect when performing the linear mixing. A base case by using traditional alternatives such as splitting the data on stationary domains is pendent as well, in order to have a way of comparison for the improvement on the estimation made by the presented methodology, if that is the case.

Finally, the possibility of investigating a statistical approach to the modeling of correlation matrices is an interesting path to follow, setting as objective to obtain a probability distribution of correlation matrices at unknown locations, improving our capabilities and understanding when modeling uncertainty, by building different scenarios sampled from such distributions.

\begin{acknowledgements}
	The author acknowledge the funding provided by the International Association for Mathematical Geosciences (IAMG) student grant, funding reference number MG-2020-14, and by the Natural Sciences and Engineering Council of Canada (NSERC), funding reference number RGPIN-2017-04200 and RGPAS-2017-507956.
\end{acknowledgements}

% Authors must disclose all relationships or interests that 
% could have direct or potential influence or impart bias on 
% the work: 
%
% \section*{Conflict of interest}
%
% The authors declare that they have no conflict of interest.

% BibTeX users please use one of
\bibliographystyle{spbasic}      % basic style, author-year citations
\bibliography{LVCM}   % name your BibTeX data base
%
% Non-BibTeX users please use
%\begin{thebibliography}{}
%%
%% and use \bibitem to create references. Consult the Instructions
%% for authors for reference list style.
%%
%\bibitem{RefJ}
%% Format for Journal Reference
%Author, Article title, Journal, Volume, page numbers (year)
%% Format for books
%\bibitem{RefB}
%Author, Book title, page numbers. Publisher, place (year)
%% etc
%\end{thebibliography}

\end{document}